\documentclass[fleqn,usenatbib]{mnras}

\usepackage{newtxtext,newtxmath}

\usepackage[T1]{fontenc}

\DeclareRobustCommand{\VAN}[3]{#2}
\let\VANthebibliography\thebibliography
\def\thebibliography{\DeclareRobustCommand{\VAN}[3]{##3}\VANthebibliography}

\usepackage{graphicx}	
\usepackage{amsmath}	
\usepackage{multirow}
\usepackage{ragged2e}
\usepackage{pdflscape}
\usepackage[flushleft]{threeparttable}
\usepackage{caption}

\newcommand{\cii}{[C\,{\sc ii}]}

\newcommand{\oiii}{[O\,{\sc iii}]}

\newcommand{\oiiil}{[O\,{\sc iii}] 88\,$\mu{\rm m}$}
\newcommand{\ciil}{[C\,{\sc ii}] 158\,$\mu{\rm m}$}

\title[An accurately constrained dust SED at $z=7.31$]{Accurate Simultaneous Constraints on the Dust Mass, Temperature and Emissivity Index of a Galaxy at Redshift 7.31}

\author[H.S.B. Algera et al.]{
Hiddo S. B. Algera,$^{1,2}$\thanks{E-mail: algera@hiroshima-u.ac.jp}
Hanae Inami,$^{1}$
Ilse De Looze,$^{3}$
Andrea Ferrara,$^{4}$
Hiroyuki Hirashita,$^{5,6}$
\newauthor
Manuel Aravena,$^{7}$
Tom Bakx,$^{8}$ 
Rychard Bouwens,$^{9}$ 
Rebecca A. A. Bowler,$^{10}$ 
Elisabete Da Cunha,$^{11,12}$
\newauthor
Pratika Dayal,$^{13}$ 
Yoshinobu Fudamoto,$^{14,2}$
Jacqueline Hodge,$^{9}$ 
Alexander Hygate,$^{9}$ 
Ivana van Leeuwen,$^{9}$
\newauthor
Themiya Nanayakkara,$^{15}$ 
Marco Palla,$^{3,16}$
Andrea Pallottini,$^{4}$
Lucie Rowland,$^{9}$ 
Renske Smit,$^{17}$
\newauthor
Laura Sommovigo,$^{18}$
Mauro Stefanon,$^{19,20}$ 
Aswin P.\ Vijayan,$^{21,22}$ 
and Paul van der Werf$^{9}$ \\
$^{1}$Hiroshima Astrophysical Science Center, Hiroshima University, 1-3-1 Kagamiyama, Higashi-Hiroshima, Hiroshima 739-8526, Japan \\
$^{2}$National Astronomical Observatory of Japan, 2-21-1, Osawa, Mitaka, Tokyo, Japan \\
$^{3}$Sterrenkundig Observatorium, Ghent University, Krijgslaan 281 - S9, 9000 Gent, Belgium \\
$^{4}$Scuola Normale Superiore, Piazza dei Cavalieri 7, I-56126 Pisa, Italy \\
$^{5}$Institute of Astronomy and Astrophysics, Academia Sinica, Astronomy-Mathematics Building, No. 1, Section 4, Roosevelt Road, Taipei 10617, Taiwan \\
$^{6}$Theoretical Astrophysics, Department of Earth and Space Science, Osaka University, 1-1 Machikaneyama, Toyonaka, Osaka 560-0043, Japan \\
$^{7}$Instituto de Estudios Astrof\'{\i}sicos, Facultad de Ingenier\'{\i}a y Ciencias, Universidad Diego Portales, Av. Ej\'ercito 441, Santiago, Chile \\
$^{8}$Department of Space, Earth, \& Environment, Chalmers University of Technology, Chalmersplatsen 4, SE-412 96 Gothenburg, Sweden \\
$^{9}$Leiden Observatory, Leiden University, NL-2300 RA Leiden, Netherlands \\
$^{10}$Jodrell Bank Centre for Astrophysics, Department of Physics and Astronomy, School of Natural Sciences, The University of Manchester, Manchester M13 9PL, UK \\
$^{11}$International Centre for Radio Astronomy Research, University of Western Australia, 35 Stirling Hwy, Crawley, 26WA 6009, Australia \\
$^{12}$ARC Centre of Excellence for All Sky Astrophysics in 3 Dimensions (ASTRO 3D) \\
$^{13}$Kapteyn Astronomical Institute, University of Groningen, P.O. Box 800, 9700 AV Groningen, The Netherlands \\
$^{14}$Waseda Research Institute for Science and Engineering, Faculty of Science and Engineering, Waseda University, 3-4-1 Okubo,
Shinjuku, Tokyo 169-8555, Japan \\
$^{15}$Centre for Astrophysics and Supercomputing, Swinburne University of Technology, PO Box 218, Hawthorn, VIC 3122, Australia \\
$^{16}$INAF - OAS, Osservatorio di Astrofisica e Scienza dello Spazio di Bologna, via Gobetti 93/3, 40129 Bologna, Italy \\
$^{17}$Astrophysics Research Institute, Liverpool John Moores University, 146 Brownlow Hill, Liverpool L3 5RF, UK \\
$^{18}$Center for Computational Astrophysics, Flatiron Institute, 162 5th Avenue, New York, NY 10010, USA \\
$^{19}$Departament d'Astronomia i Astrof\`isica, Universitat de Val\`encia, C. Dr. Moliner 50, E-46100 Burjassot, Val\`encia, Spain \\
$^{20}$Unidad Asociada CSIC "Grupo de Astrof\'isica Extragal\'actica y Cosmolog\'ia" (Instituto de F\'isica de Cantabria - Universitat de Val\`encia) \\
$^{21}$Cosmic Dawn Center (DAWN) \\
$^{22}$DTU-Space, Technical University of Denmark, Elektrovej 327, DK-2800 Kgs. Lyngby, Denmark
}

\date{Accepted XXX. Received YYY; in original form ZZZ}

\pubyear{2024}

\begin{document}
\label{firstpage}
\pagerange{\pageref{firstpage}--\pageref{lastpage}}
\maketitle

\begin{abstract}
We present new multi-frequency ALMA continuum observations of the massive [$\log_{10}(M_\star/M_\odot) = 10.3_{-0.2}^{+0.1}$], UV-luminous [$M_\mathrm{UV} = -21.7 \pm 0.2$] $z=7.31$ galaxy REBELS-25 in Bands 3, 4, 5, and 9. Combining the new observations with previously-taken data in Bands 6 and 8, we cover the dust continuum emission of the galaxy in six distinct bands -- spanning rest-frame $50-350\,\mu$m -- enabling simultaneous constraints on its dust mass ($M_\mathrm{dust}$), temperature ($T_\mathrm{dust}$) and emissivity index ($\beta_\mathrm{IR}$) via modified blackbody fitting. Given a fiducial model of optically thin emission, we infer a cold dust temperature of $T_\mathrm{dust} = 32_{-6}^{+9}\,$K and a high dust mass of $\log_{10}(M_\mathrm{dust}/M_\odot) = 8.2_{-0.4}^{+0.6}$, and moderately optically thick dust does not significantly alter these estimates. If we assume dust production is solely through supernovae (SNe), the inferred dust yield would be high, $y = 0.7_{-0.4}^{+2.3}\,M_\odot$ per SN. Consequently, we argue grain growth in the interstellar medium of REBELS-25 also contributes to its dust build-up. This is supported by the steep dust emissivity index $\beta_\mathrm{IR} = 2.5 \pm 0.4$ we measure for REBELS-25, as well as by its high stellar mass, dense interstellar medium, and metal-rich nature. Our results suggest that constraining the dust emissivity indices of high-redshift galaxies is important not only to mitigate systematic uncertainties in their dust masses and obscured star formation rates, but also to assess if dust properties evolve across cosmic time. We present an efficient observing setup to do so with ALMA, combining observations of the peak and Rayleigh-Jeans tail of the dust emission.
\end{abstract}




\section{Introduction}
\label{sec:introduction}

A thorough understanding of both the formation and properties of cosmic dust at high redshift is crucial for our understanding of the evolution of galaxies as a whole. The presence of dust can be indirectly inferred by observing the attenuated rest-frame UV and optical emission of galaxies, or directly through observing its signature infrared emission (e.g., \citealt{galliano2018}). Besides shaping the spectral energy distribution (SED) of galaxies, dust also catalyzes the formation of molecular hydrogen from which new stars can subsequently form \citep{gould1963}.

In recent years, observations with the Atacama Large Millimeter/submillimeter Array (ALMA) have demonstrated the rapid emergence of dust in the early Universe (\citealt{watson2015,bowler2018,hashimoto2019,tamura2019,bakx2021,schouws2022,inami2022,witstok2022}). In particular, submillimeter observations of relatively massive galaxies ($M_\star \approx 10^{8.5-10}\,M_\odot$) at $4\lesssim z \lesssim 7$ have revealed that, on average, nearly half of their star formation is hidden behind dust \citep{bowler2018,bowler2023,fudamoto2020,inami2022,algera2023,mitsuhashi2023a,mitsuhashi2023b}, and that even at $z\sim7$ up to $\sim30\,\%$ of the cosmic star formation rate density (SFRD) may be dust-obscured \citep{algera2023,barrufet2023}.

However, the bulk of the dust detections in the epoch of reionization ($z\gtrsim6.5$) remain limited to a single continuum measurement (e.g., \citealt{inami2022}). While such observations establish the presence of dust, single-band measurements are insufficient to characterize its physical properties (e.g., \citealt{bakx2021,algera2024}). Multi-band continuum observations are crucial to reveal dust masses ($M_\mathrm{dust}$), temperatures ($T_\mathrm{dust}$) and -- if more than two bands are available -- emissivity indices ($\beta_\mathrm{IR}$). Given that the infrared-based star formation rate (SFR) scales as $\mathrm{SFR}_\mathrm{IR} \propto M_\mathrm{dust} T_\mathrm{dust}^{\beta_\mathrm{IR}+4}$, obscured star formation rates can therefore only be accurately determined when these parameters are robustly constrained.

Dust temperature measurements of high-redshift galaxies have garnered significant attention in recent years, both from an observational and theoretical perspective (e.g., \citealt{behrens2018,schreiber2018,liang2019,faisst2020,sommovigo2020,sommovigo2021,sommovigo2022,sommovigo2022b,bakx2020,bakx2021,bakx2024,hirashita2022,vijayan2022,witstok2022,fudamoto2023,mauerhofer2023,algera2024}). Most studies now agree that dust is typically warmer in early galaxies (e.g., \citealt{reuter2020,sommovigo2022}), although the extent to which average dust temperatures evolve remains unclear. In particular, the scatter among dust temperatures of high-redshift galaxies is enormous, suggesting a diverse underlying population (e.g., \citealt{bakx2020,witstok2022,algera2024}).

When robust dust temperature measurements are available, galaxy dust masses can simultaneously be accurately constrained, and subsequently compared to theoretical models of dust assembly. Core-collapse supernovae (CC-SNe) are typically considered the main dust producers in the early Universe (e.g., \citealt{todini2001,michalowski2015,vijayan2019,dayal2022}). However, the precise dust yields per SN remain debated in the literature, and may be too low to fully explain observed dust masses at high redshift \citep{bianchi2007,lesniewska2019,slavin2020}, which can be as high as $M_\mathrm{dust} \approx 10^{8-9}\,\mathrm{M}_\odot$ in particularly massive galaxies or quasars at $z\gtrsim6$ \citep{venemans2018,wang2021,witstok2022,decarli2023,algera2024,tripodi2024}. 

To explain such high dust masses, grain growth in the ISM of distant galaxies is often invoked (e.g., \citealt{dicesare2023}). However, efficient grain growth likely relies upon the galaxy already being relatively metal-enriched \citep{asano2013}, which requires early episodes of star formation. In addition, a dense ISM is likely required for grain growth to proceed efficiently \citep{hirashita2012,ferrara2016}. Furthermore, while dust produced by AGB stars may contribute to the overall dust budget as well, it is likely subdominant at $z\gtrsim6$ given the longer timescales involved to reach this late stellar phase (\citealt{valiante2011,dayal2022}; though see e.g., \citealt{sloan2009}). Finally, other stellar sources such as Wolf-Rayet stars (e.g., \citealt{lau2020}) and Red Super Giants (e.g., \citealt{levesque2006,nozawa2014}) may also produce dust, although likely not in amounts that significantly contribute to the dust masses of high-redshift galaxies (see \citealt{schneider2023} for a recent review).

The interplay between the different mechanisms of dust production in early galaxies is expected to leave its imprint on their spectral energy distributions. Recently, \citet{witstok2023a} found evidence for carbonaceous dust in the rest-optical SED of a $z=6.71$ galaxy through the detection of its characteristic $2175\,$\AA~bump feature (see also \citealt{markov2023}). Similarly, at far-infrared wavelengths the dust emissivity index $\beta_\mathrm{IR}$ encapsulates the physical properties of the underlying dust grains, such as their size distribution and composition (e.g., \citealt{galliano2018,inoue2020}) -- albeit in a non-trivial way \citep{ysard2019}. Nevertheless, accurately measuring $\beta_\mathrm{IR}$ in samples of high-redshift galaxies, and investigating whether there is any discernible evolution from dust properties observed locally, may hold clues as to the production avenues of the earliest dust.

In this paper, we present new ALMA continuum measurements of the $z=7.31$ galaxy REBELS-25, spanning a rest-frame wavelength range of $\lambda_\mathrm{rest} = 50 - 350\,\mu$m. In Section \ref{sec:observations} we discuss the newly acquired ALMA data, and we outline our framework for fitting the infrared SED of our target in Section \ref{sec:methods}. We subsequently present this SED in Section \ref{sec:results}, and discuss our findings in Section \ref{sec:discussion}. Finally, we discuss efficient means of constraining the dust SEDs of distant galaxies in Section \ref{sec:mockfitting}, before leaving the reader with our concluding remarks in Section \ref{sec:conclusions}. Throughout this work, we assume a standard $\Lambda$CDM cosmology, with $H_0=70\,\text{km\,s}^{-1}\text{\,Mpc}^{-1}$, $\Omega_m=0.30$ and $\Omega_\Lambda=0.70$. We further adopt a \citet{chabrier2003} initial mass function.

\section{Target, Observations and Flux Density Measurements}
\label{sec:observations}

\begin{table*}
    \centering
    \caption{Summary of the multi-band ALMA continuum image properties of REBELS-25.}
    \label{tab:imaging}
    \begin{tabular}{ccccccclcc}
        \hline
        \textbf{Band} & $\nu_\mathrm{cen}$ & \textbf{RMS} & $\theta_\mathrm{major}$ & $\theta_\mathrm{minor}$ & $\mathrm{PA}$ & \textbf{PID} & \textbf{PI} & \textbf{Flux Density} & \textbf{Uncertainty} \\
        \hline
         & [GHz] & [$\mu\mathrm{Jy\,bm}^{-1}$] & & & & & & [$\mu\mathrm{Jy}$] & [$\mu\mathrm{Jy}$] \\
        \hline
        3 & 102.5 & 4.0 & $0\farcs80$ & $0\farcs76$ & $+34.4^\circ$ & 2021.1.01495.S & De Looze & 16.9 & 3.9 \\ 
        4 & 145.9 & 8.2 & $1\farcs89$ & $1\farcs28$ & $-58.9^\circ$ & 2022.1.01324.S & Algera & 34.9 & 7.8 \\
        5 & 168.9 & 8.9 & $1\farcs55$ & $1\farcs02$ & $-63.4^\circ$ & 2022.1.01384.S & Fudamoto & 81.3 & 10.1 \\ 
        6 & 227.3 & 9.0 & $1\farcs06$ & $0\farcs94$ & $-86.7^\circ$ & 2017.1.01217.S, 2019.1.01634.L & Stefanon, Bouwens & 226.5 & 13.9 \\
        8 & 403.4 & 112 & $1\farcs09$ & $0\farcs64$ & $-63.4^\circ$ & 2021.1.00318.S & Inami & 641.4 & 146.9 \\
        9 & 687.6 & 276 & $0\farcs48$ & $0\farcs36$ & $+84.4^\circ$ & 2022.1.01324.S & Algera & $< 1098$ & 366 \\     \hline
    \end{tabular}
    \justify
    \textbf{Notes:} Col.\ 1: ALMA Observing band. Col.\ 2: Central observed-frame frequency. Col.\ 3: root-mean-square noise in the naturally-weighted map.  Cols.\ 4, 5, 6: major axis, minor axis and position angle of the synthesized beam in the naturally-weighted images. Cols.\ 7, 8: ALMA program ID and principal investigator. Cols.\ 9, 10: peak flux density of REBELS-25 and corresponding uncertainty, after tapering to ensure the galaxy is unresolved.
\end{table*}

Our target, REBELS-25, is a UV-luminous galaxy ($M_\mathrm{UV} = -21.7 \pm 0.2$) drawn from the Reionization Era Bright Emission Line Survey (REBELS), a Cycle 7 ALMA Large Program aimed at constraining the dust and ISM reservoirs of 40 UV-selected galaxies at $z\gtrsim6.5$ \citep{bouwens2022}. REBELS-25 was identified as the brightest galaxy within REBELS in Band 6 dust continuum emission, indicative of significant obscured star formation \citep{inami2022}. In addition, REBELS-25 was spectroscopically confirmed through its bright \ciil{} line, and has a \cii{}-based redshift of $z_\mathrm{spec} = 7.3065 \pm 0.0001$ \citep{hygate2023}. Spectral energy distribution fitting of its rest-frame UV-to-optical emission suggests REBELS-25 to be a massive galaxy, with $\log_{10}(M_\star / \mathrm{M}_\odot) = 10.3_{-0.2}^{+0.1}$ \citep{topping2022}. Moreover, \citet{rowland2024} recently investigated the kinematics of REBELS-25 using high-resolution \cii{} observations, and demonstrated that REBELS-25 is a dynamically cold disk galaxy in the Epoch of Reionization.

Follow-up Band 8 observations in ALMA Cycle 8 provided a first dual-band measurement of the dust temperature of REBELS-25, $T_\mathrm{dust} = 34 \pm 6\,$K, and dust mass $\log (M_\mathrm{dust} / M_\odot) = 8.1 \pm 0.3$, assuming a fiducial $\beta_\mathrm{IR} = 2$ \citep{algera2024}. Given this fiducial model, the total SFR of REBELS-25 -- summing the contributions from unobscured and obscured star formation -- is $\mathrm{SFR}_\mathrm{tot} = 78_{-22}^{+39}\,M_\odot\,\mathrm{yr}^{-1}$ \citep{hygate2023,algera2024}.

\begin{figure}
    \centering
    \includegraphics[width=0.5\textwidth]{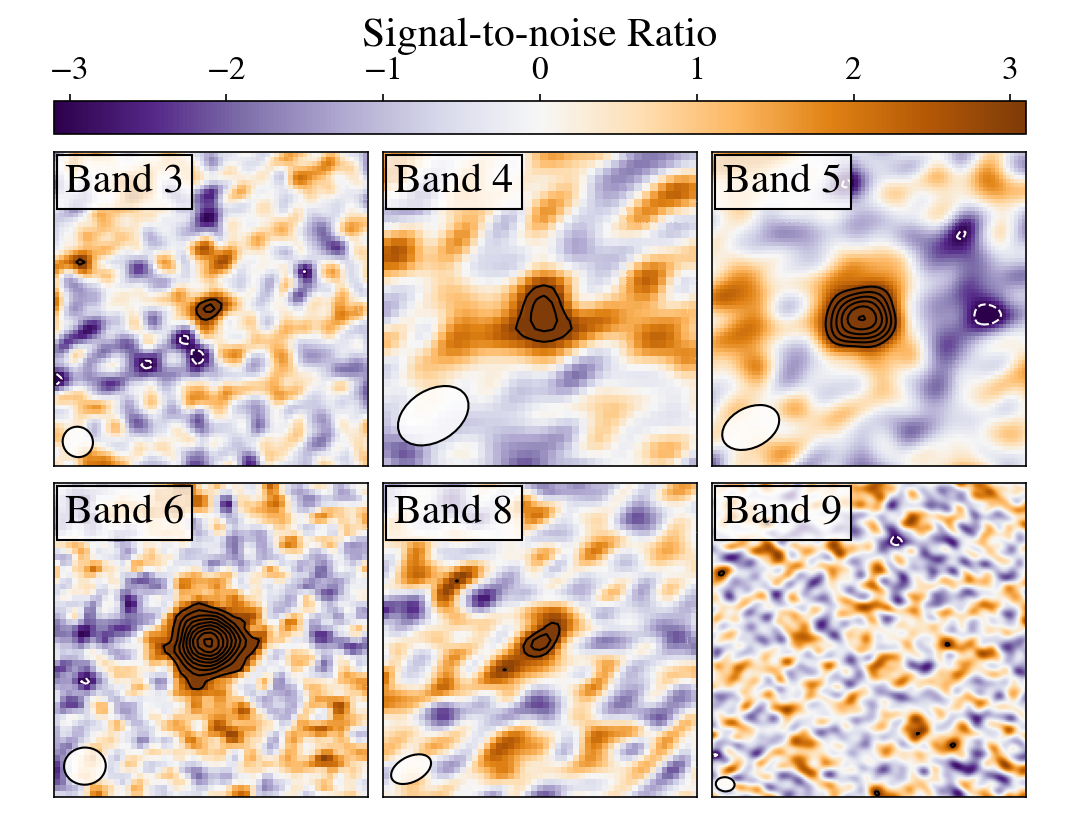}
    \caption{Cutouts ($8''\times8''$) centered on the dust continuum emission from REBELS-25 in six ALMA bands. All images are shown using natural weighting, with their respective beam sizes indicated in the lower left corner via the white ellipse. Black contours represent positive emission, and start at $3\sigma$ and increase in steps of $1\sigma$ for all bands except Band 6, where contours instead increase in steps of $2\sigma$ to aid visual clarity. Here $\sigma$ is the RMS in the image. Negative emission is shown through white, dashed contours. Dust emission from REBELS-25 is detected in all bands but Band 9.}
    \label{fig:cutouts}
\end{figure}

In this work, we present newly obtained Cycle 9 ALMA observations of REBELS-25 in Bands 3, 4, 5 and 9 to perform a simultaneous measurement of the dust mass, temperature and emissivity index of this massive star-forming galaxy at $z=7.31$. We made use of the ALMA pipeline in {\sc{casa}} (Common Astronomy Software Applications; \citealt{casa2022}) to calibrate the newly acquired Cycle 9 data, adopting the pipeline version stated in the corresponding QA2 reports. The Band 4 observations required some additional a priori flagging, while for the remaining bands the standard pipeline calibrations were used. We initially imaged the data via task {\sc{tclean}} using natural weighting to optimize the continuum sensitivity, adopting a cleaning threshold of $2\times$ the root-mean-square (RMS) noise.\footnote{A threshold of $1-3\sigma$ is typical (see e.g., \citealt{algera2021,garcia-vergara2022,jones2023,posses2024}; see also the detailed discussion in \citealt{czekala2021}).} In this analysis, we furthermore ensured that the frequencies within $2\times$ the full-width-half-maximum (FWHM) of the \ciil{} \citep{hygate2023}, \oiiil{} \citep{algera2024} and CO(7-6) (A.\ Hygate et al.\ in prep) lines were excluded. No other emission lines are detected in the data. We summarize the image parameters in Table \ref{tab:imaging} and present the six-band, naturally-weighted dust continuum maps of REBELS-25 in Figure \ref{fig:cutouts}. The dust continuum is clearly detected in Bands 3, 4, 5, 6 and 8, while it remains undetected in Band 9 at the $3\sigma$ level.

Given the variation in the native beam sizes of our multi-frequency ALMA data, we re-image the observations by applying $uv$-tapering to ensure consistent flux density measurements. The effect of tapering is to reduce the resolution of the images by down-weighting long-baseline data. As a result, once tapered to a sufficiently coarse resolution, the galaxy will become unresolved and its total flux density is well-represented by the peak flux density. On the other hand, the RMS in the image increases for coarser $uv$-tapers as a fraction of the data is effectively discarded. We taper each of the bands to a variety of coarser beams, ranging from the native beam size to $2\farcs0$ or $3\farcs0$ depending on the original resolution, and measure the peak and integrated flux densities in each of the tapered images through 2D Gaussian fitting with {\sc{PyBDSF}} \citep{mohanrafferty2015}. We verify that, at sufficiently coarse $uv$-tapers, the peak and integrated flux densities agree within the uncertainties. Moreover, we note that, compared to integrated fluxes, peak flux densities tend to be more robust against the inclusion of noise peaks when the signal-to-noise ratio of the data is modest (e.g., \citealt{vandervlugt2021}). 

We choose to adopt the peak flux density at the lowest tapering (that is, in the highest resolution image) where the flux measurement converges to a constant value, signifying the source is unresolved at this resolution. In Band 3, the measured peak (and integrated) flux density does not change appreciably depending on the applied $uv$-taper, and as such we adopt the peak flux density measured in the naturally-weighted map. In Bands 4, 5 and 8, we adopt a $uv$-taper ranging from $1\farcs4 - 1\farcs6$.\footnote{Given the asymmetric beam in Band 4 (Table \ref{tab:imaging}), the final $uv$-tapering we adopt is only applied along the minor beam axis.} For Band 6, where REBELS-25 is detected at high S/N and was previously found to be resolved at the native $\sim1\farcs0$ resolution \citep{inami2022}, we adopt a coarser tapering of $2\farcs5$. The uncertainty on the peak flux density is determined through {\sc{PyBDSF}}, and approximately equals the RMS in the corresponding tapered map. For Band 9, where REBELS-25 is not detected, we conservatively quote the upper limit as $3$ times the RMS noise in the $0\farcs8$ tapered map. This choice is a trade-off between degrading the resolution to ensure the (undetected) emission from REBELS-25 is not spread across multiple beams, and retaining adequate sensitivity to quote a meaningful upper limit. In practice, this upper limit is $\sim30\,\%$ larger (i.e., more conservative) than would be obtained from the naturally-weighted map. The ALMA flux densities of REBELS-25 are presented in Table \ref{tab:imaging}. We note that our results do not change when we adopt a common UV-taper $\gtrsim1\farcs5$ across all bands. Furthermore, we find consistent results when performing aperture photometry on either the naturally-weighted or tapered images.

\section{Methods}
\label{sec:methods}

We adopt the Modified Blackbody (MBB) fitting framework introduced in \citet{algera2024} to fit the dust SED of REBELS-25. We follow \citet{dacunha2013} by including a correction for the cosmic microwave background (CMB), which both forms a source of heating of the dust, and a background against which it is observed. The explicit equation used in the fitting is (reproduced from \citealt{algera2024}):

{\footnotesize
\begin{align}
    S_\nu = \left(\frac{1+z}{d_\mathrm{L}^2}\right) \left(\frac{1 - e^{-\tau_\nu}}{\tau_\nu}\right) M_\mathrm{dust} \kappa_0 \left(\frac{\nu}{\nu_0}\right)^{\beta_\mathrm{IR}} \left[B_\nu(T_{\mathrm{dust},z}) - B_\nu(T_{\mathrm{CMB},z}) \right] \ .
    \label{eq:mbb_fit}
\end{align}
}

\noindent This expression contains four free parameters: the dust temperature $T_\mathrm{dust}$,\footnote{We distinguish between the dust temperature including CMB-heating, $T_{\mathrm{dust},z}$ and the temperature corrected for the CMB, $T_\mathrm{dust}$. All quoted far-IR luminosities are corrected for CMB-heating, and hence reflect the emission due to obscured star formation.} dust mass $M_\mathrm{dust}$, emissivity index $\beta_\mathrm{IR}$, and optical depth $\tau_\nu$ at rest-frame frequency $\nu$. The optical depth can be written as $\tau_\nu = (\nu / \nu_\mathrm{thick})^{\beta_\mathrm{IR}} = (\lambda / \lambda_\mathrm{thick})^{-\beta_\mathrm{IR}}$, where $\lambda_\mathrm{thick} = c / \nu_\mathrm{thick}$ is the rest-frame wavelength where the dust transitions from optically thin to optically thick (that is, $\tau_\nu(\lambda_\mathrm{thick}) = 1$). We adopt a fixed dust opacity $\kappa_0 = \kappa(\nu_0=1900\,\mathrm{GHz}) = 10.41\,\mathrm{cm}^2\,\mathrm{g}^{-1}$, corresponding to Milky Way dust \citep{weingartner2001}. While the true dust opacity depends on its chemical composition -- which is inherently unknown at high redshift -- this value has been adopted in numerous recent works (e.g., \citealt{ferrara2022,schouws2022,sommovigo2022,algera2024}), enabling a direct comparison. Finally, we define the luminosity distance at redshift $z$ as $d_\mathrm{L}$ and write $B_\nu$ to denote the Planck function in frequency units. 

We use a Monte Carlo Markov Chain (MCMC) technique to fit the continuum flux densities of REBELS-25, which ensures the possible parameter space is accurately sampled, and thus provides robust uncertainties on the derived parameters (e.g., \citealt{foreman-mackey2013}). We adopt a flat prior on the dust mass of $\log (M_\mathrm{dust} / M_\odot) \in [4,12]$, and a smooth prior on the dust temperature combining a flat prior between $T_{\mathrm{dust},z} = [T_{\mathrm{CMB},z}, 150]\,$K, and a smooth decline at temperatures above $150\,$K via a Gaussian of $\sigma=30\,$K \citep{algera2024}. At the redshift of REBELS-25, $T_\mathrm{CMB}(z=7.31) = 22.6\,$K. Finally, we adopt a flat and purposefully wide prior on the emissivity index of $\beta_\mathrm{IR}\in[1,4]$ (see Section \ref{sec:discussion_mockfitting}). We emphasize that the precise priors adopted do not significantly affect the inferred dust parameters for REBELS-25, given that its SED is accurately constrained across multiple bands.

At this stage, it is furthermore important to note that dust temperatures obtained from MBB-fitting do not necessarily constitute an accurate representation of any physical dust temperature within the galaxy (e.g., \citealt{casey2012,liang2019,lower2024}). In reality, dust of various temperatures will be present throughout the ISM of any given galaxy, resulting in a dust temperature distribution that can be characterized by either a mass- or luminosity-weighted temperature. In practice, the temperature obtained from MBB-fitting typically falls in between these two extremes (\citealt{lower2024}; see also L.\ Sommovigo \& H.\ Algera et al.\ in prep), although it is further affected by optical depth effects. In the case of optically thin dust, the dust mass of a REBELS-25-like galaxy can be accurately inferred from MBB-fitting despite the simplifying assumption of a single dust temperature: the dust mass might be marginally underestimated by $\sim0.1-0.2\,$dex, but this is well within the current measurement uncertainties (see below, and L.\ Sommovigo \& H.\ Algera et al.\ in prep).

\section{Results}
\label{sec:results}

We next fit the dust continuum SED of REBELS-25, and investigate to what extent the recovered dust parameters depend upon the assumption of optically thin or optically thick dust emission.

\subsection{Optically Thin Dust}
\label{sec:results_thin}

\begin{table}
    \def\arraystretch{1.15}
    \centering
    \caption{Dust properties inferred from modified blackbody fitting to the ALMA continuum photometry of REBELS-25.}
    \label{tab:fluxes}
    \begin{tabular}{llll}
        \hline
         \textbf{Parameter} & \textbf{Opt.\ Thin} & \textbf{Fixed $\mathbf{\lambda_\mathrm{thick}}$} & \textbf{Free $\mathbf{\lambda_\mathrm{thick}}$} \\
        \hline 
        $\lambda_\mathrm{thick}\,[\mu\mathrm{m}]$ & 0 & 65 & $152_{-72}^{+41}$ \\
        $\log_{10}(M_\mathrm{dust} / \mathrm{M}_\odot)$ & $8.2_{-0.4}^{+0.6}$ & $8.0_{-0.4}^{+0.6}$ & $7.4_{-0.3}^{+0.5}$\\
        $T_\mathrm{dust}\,[\mathrm{K}]$ & $32_{-6}^{+9}$ & $37_{-9}^{+13}$ & $73_{-34}^{+55} $\\
        $\beta_\mathrm{IR}$ & $2.5 \pm 0.4$ & $2.5_{-0.4}^{+0.5}$ & $2.7 \pm 0.5$ \\
        $\log_{10}(L_\mathrm{IR} / \mathrm{L}_\odot)$ & $11.7_{-0.2}^{+0.3}$ & $11.7_{-0.1}^{+0.2}$ & $12.0_{-0.3}^{+0.4} $\\ 
        $\log_{10}(M_\mathrm{dust} / M_\star)^\dagger$ & $-2.1_{-0.4}^{+0.6}$ & $-2.2_{-0.4}^{+0.6}$ & $-2.8_{-0.4}^{+0.6}$ \\
        \hline 
        \end{tabular}
    \raggedright \justify \vspace{-0.2cm}
    \textbf{Notes:} Col.\ 2: Optically thin MBB fit in Sec.\ \ref{sec:results_thin}. Col.\ 3: Optically thick MBB fit with $\lambda_\mathrm{thick} = 65\,\mu$m in Sec.\ \ref{sec:results_thick}. Col.\ 4: Optically thick MBB fit with $\lambda_\mathrm{thick}$ as a free parameter in Sec.\ \ref{sec:results_vary_thick}.
    $\dagger$ Dust-to-stellar-mass ratio, assuming the stellar mass from \citet{topping2022}.
\end{table}

\begin{figure}
    \centering
    \hspace*{-0.45cm}\includegraphics[width=0.53\textwidth]{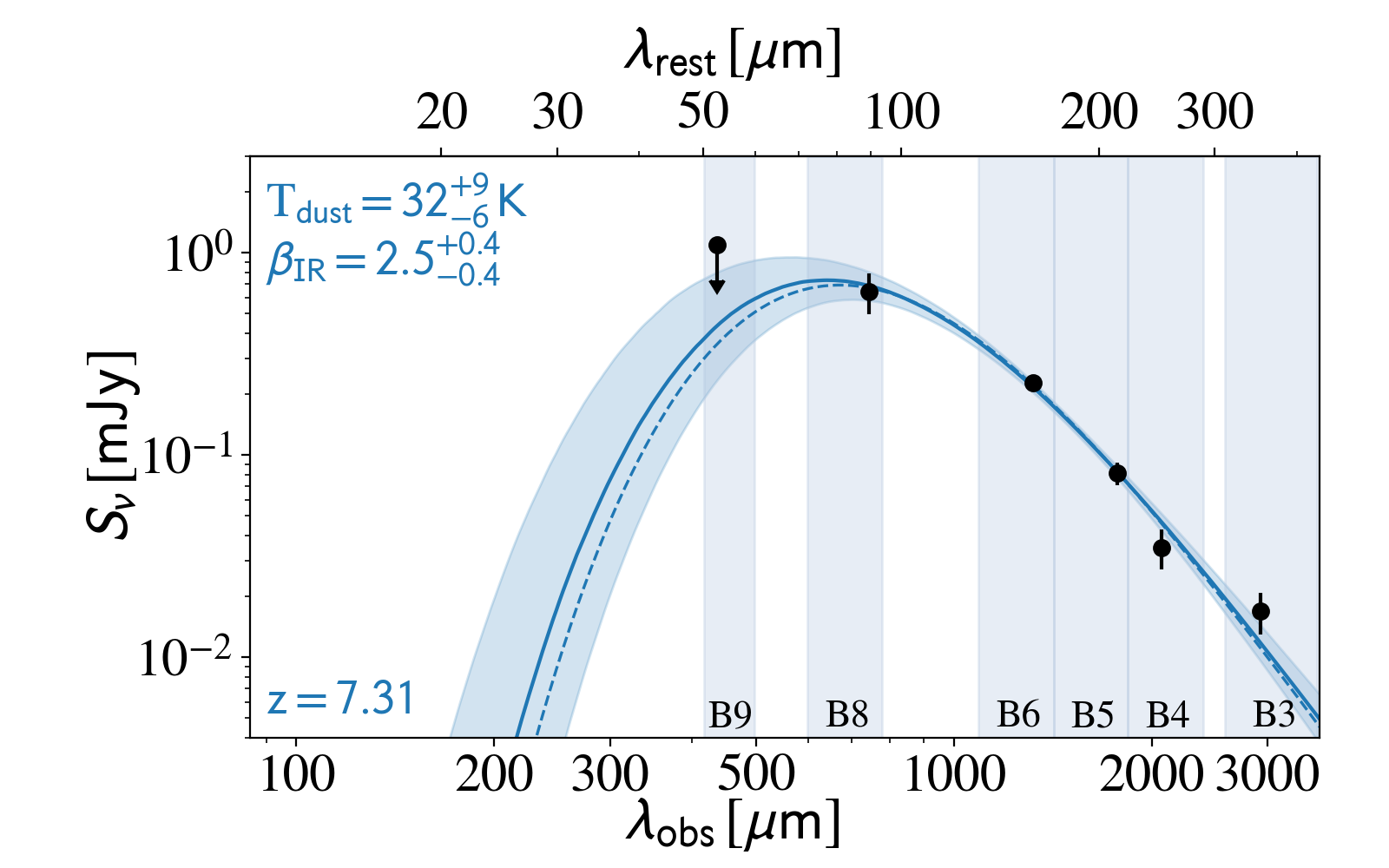}
    \caption{Optically thin modified blackbody fit to the multi-band ALMA observations of the $z=7.31$ galaxy REBELS-25. Owing to the broad coverage of its far-infrared SED, the dust mass, temperature and emissivity index can be accurately simultaneously constrained. The solid line and blue shading show the median and corresponding $16 - 84^\mathrm{th}$ percentiles of the fit, while the dashed line shows the fit that maximizes the log-likelihood (maximum a posteriori solution). The various ALMA bands in which REBELS-25 has been observed are indicated via the vertical shading. The fit demonstrates that REBELS-25 hosts a cold dust reservoir, with a steep dust emissivity index.}
    \label{fig:MBBfits}
\end{figure}

We first assume the dust SED of REBELS-25 can be described by a fully optically thin modified blackbody. We show the resulting six-band fit in Figure \ref{fig:MBBfits}. Assuming optically thin dust, we infer a dust temperature of $T_\mathrm{dust} = 32_{-6}^{+9}\,$K, confirming the previously measured cold dust temperature for REBELS-25 by \citet{algera2024}, who also adopted optically thin dust, but with a fixed $\beta_\mathrm{IR} = 2$.\footnote{Throughout this work, we use `cold' in the context of the known high-redshift galaxy population. For example, nearby galaxies detected in \textit{Herschel} observations tend to have dust temperatures of $T_\mathrm{dust} \sim 20 - 30\,$K \citep{cortese2014,lamperti2019}, below that inferred for REBELS-25. However, both the presence of a temperature floor in the CMB and the typically more star-forming conditions in galaxies at high redshift suggest these, on average, have higher dust temperatures than local systems ($T_\mathrm{dust} \gtrsim 40 - 50\,$K; e.g., \citealt{bethermin2015,schreiber2018,sommovigo2022}).} Intriguingly, this is only $\sim10\,$K warmer than the CMB at $z=7.31$. Despite now also varying $\beta_\mathrm{IR}$ in the fit, the uncertainty on the recovered dust temperature is similar to that in the dual-band fit from \citet{algera2024}.

We additionally measure a high dust mass for REBELS-25, of $\log_{10}(M_\mathrm{dust} / M_\odot) = 8.2_{-0.4}^{+0.6}$, as well as a dust emissivity index $\beta_\mathrm{IR} = 2.5 \pm 0.4$. As we will discuss in Section \ref{sec:discussion}, this value of $\beta_\mathrm{IR}$ appears steeper than the typically assumed value of $\beta_\mathrm{IR} \sim 1.5 - 2$ (e.g., \citealt{galliano2018}). Utilizing the full posterior distribution for $\beta_\mathrm{IR}$, we find the probability of $\beta_\mathrm{IR} \leq 2$ amounts to approximately $13\%$. The full set of recovered fitting parameters is presented in Table \ref{tab:fluxes}.

We moreover note that the Rayleigh-Jeans tail of the dust SED in Fig.\ \ref{fig:MBBfits} shows minor deviations around the best fit. In Bands 3 and 4, the dust emission is detected only at the $4.3 - 4.5\sigma$ level, so this is likely simply due to noise. We note that REBELS-25 is not detected in the $3\,$GHz Very Large Array observations covering the COSMOS field down to $\sigma = 2.3\,\mu\mathrm{Jy\,beam}^{-1}$ \citep{smolcic2017}, which rules out a significant contribution from non-thermal synchrotron emission to the Band 3 flux density. Moreover, as these observations probe a rest-frame wavelength of $\sim0.35\,\mathrm{mm}$, we furthermore do not expect any contribution from thermal free-free emission (e.g., \citealt{algera2021,algera2022}). We note that, if we omit either the Band 3 or Band 4 datapoint from the MBB fit, we recover a dust emissivity index that is consistent with the fiducial value of $\beta_\mathrm{IR} = 2.5 \pm 0.4$ within the uncertainties (c.f., $\beta_\mathrm{IR,no\,B3} = 2.9_{-0.5}^{+0.4}$ and $\beta_\mathrm{IR,no\,B4} = 2.1_{-0.4}^{+0.5}$).

\subsection{Optically Thick Dust}

The high apparent dust mass and cold dust temperature of REBELS-25 inferred from optically thin models may suggest its dust emission to, in fact, be optically thick. To investigate this possibility, we re-fit the dust SED of REBELS-25 with either a fixed transition wavelength $\lambda_\mathrm{thick}$ where $\tau_\nu = 1$ (Section \ref{sec:results_thick}), or by freely varying this wavelength during the fitting process (Section \ref{sec:results_vary_thick}).

\subsubsection{A fixed transition wavelength $\lambda_\mathrm{thick}$}
\label{sec:results_thick}

We start by estimating (an upper limit on) the wavelength where the dust turns optically thick, assuming a spherical dust distribution with radius $R$. In this case, the transition wavelength $\lambda_\mathrm{thick}$ is related to the dust mass, size, and emissivity index via (e.g., \citealt{inoue2020}):

\begin{align}
    \lambda_\mathrm{thick} &= \lambda_0 \left(\frac{\pi R^2}{\kappa_0 M_\mathrm{dust}}\right)^{-\frac{1}{\beta_\mathrm{IR}}} \\
    &\approx 157.8\,\mu\mathrm{m} \times \left[ 14.5 \times \left(\frac{R}{1\,\mathrm{kpc}}\right)^{2} \left(\frac{M_\mathrm{dust}}{10^8\,\mathrm{M}_\odot}\right)^{-1}\right]^{-\frac{1}{\beta_\mathrm{IR}}}\, ,
    \label{eq:lambda_thick}
\end{align}

\noindent where $\lambda_0 = (c / \nu_0)$. In previous studies (\citealt{inami2022,hygate2023}), the Band 6 dust continuum emission of REBELS-25 was found to have a half-light radius $R_{1/2} \approx 1.5\,\mathrm{kpc}$ using the same data also used in this work.\footnote{We quote circularized radii, defined as $0.5 \times \sqrt{ab}$ where $a$ ($b$) is the major (minor) axis of a 2D Gaussian, after deconvolution from the beam.} However, the new high-resolution ($\sim0\farcs15)$ Band 6 observations of REBELS-25 recently presented in \citet{rowland2024} show that its dust continuum morphology is elongated in the north-south direction, with potential dusty clumps at the $\sim3-4\sigma$ level. We therefore utilize these new observations to infer the dust continuum size of REBELS-25, while referring to \citet{rowland2024} and H.\ Algera et al.\ (in prep) for further details. Briefly, we fit a 2D Gaussian to the dust continuum emission via {\sc{CASA}} imfit (see also \citealt{hygate2023}), either 1) including the full extent of REBELS-25, or 2) excluding the northern- and southernmost clumps. These fits yield values of $R_{1/2} = 1.13 \pm 0.10\,\mathrm{kpc}$ and $R_{1/2} = 0.87 \pm 0.08\,\mathrm{kpc}$, respectively. In what follows, we therefore adopt a fiducial dust half-light radius of $R_{1/2} \approx 1.0\,\mathrm{kpc}$ to estimate $\lambda_\mathrm{thick}$.

In applying Equation \ref{eq:lambda_thick}, we further need to adopt a dust mass for REBELS-25, for which we assume the value obtained from the optically thin MBB fit. In general, fits allowing for optically thick emission find lower dust masses than would be inferred from optically thin models (e.g., \citealt{algera2024}). This is effectively due to the peak of the dust SED being determined by both the temperature of the dust and its optical depth. Under an optically thin assumption, a shift of the peak towards longer wavelengths requires colder temperatures. However, in optically thick models, both a colder temperature and a higher optical depth can produce such a shift. In practice, this results in optically thick models recovering higher dust temperatures -- and consequently lower dust masses -- than optically thin ones. In our estimate of $\lambda_\mathrm{thick}$, we can thus safely adopt the dust mass inferred from optically thin models as a strict upper limit. Further adopting the measured $\beta_\mathrm{IR} = 2.5$, we infer $\lambda_\mathrm{thick} \lesssim 65\,\mu$m. This would imply the emission in Band 9 to be subject to optical depth effects, while the emission in Band 8 is mostly optically thin ($\tau_\nu(88\,\mu\mathrm{m}) \approx 0.5$).

To investigate the effect of such moderately optically thick dust on the derived dust parameters, we fit the dust SED of REBELS-25 assuming a fixed $\lambda_\mathrm{thick} = 65\,\mu$m. We present the corresponding fitting parameters in Table \ref{tab:fluxes} and show the fit in Figure \ref{fig:MBB_thick} in Appendix \ref{app:thick}. Under these assumptions, we recover a slightly lower -- albeit still large -- dust mass of $\log_{10}\left(M_\mathrm{dust} / \mathrm{M}_\odot\right) = 8.0_{-0.4}^{+0.6}$. However, both the dust mass and other fitting parameters are consistent with the equivalent parameters obtained from optically thin models within $1\sigma$. As such, we find that moderately optically thick dust does not appreciably alter the inferred dust parameters for REBELS-25. 

\subsubsection{A varying transition wavelength $\lambda_\mathrm{thick}$}
\label{sec:results_vary_thick}

While the analysis in the previous section suggests the recovered dust parameters of REBELS-25 are not strongly affected by optical depth effects, we caution that this estimate relies on the assumption of a spherical and homogeneous dust distribution. As this is unlikely to be the case in practice, we perform an additional fit to the dust SED of REBELS-25, this time allowing $\lambda_\mathrm{thick}$ to vary freely. We adopt a flat prior of $\lambda_\mathrm{thick} \in [0, 300]\,\mu$m, motivated by observations of submillimeter galaxies at lower redshifts, which show typical values of $\lambda_\mathrm{thick}\sim75 - 200\,\mu$m (e.g., \citealt{simpson2017,cortzen2020}).

We show the MBB fit to REBELS-25 in Figure \ref{fig:MBB_thick}. As expected, the fit is unconstrained at the shortest wavelengths where we rely on just an upper limit in ALMA Band 9. In turn, the inferred dust mass and temperature are highly uncertain (Table \ref{tab:fluxes}). The median transition wavelength is found to be $\lambda_\mathrm{thick} = 152_{-72}^{+41}\,\mu$m, indicating it is possible that optical depth effects are more significant than expected based on simple arguments assuming a spherical dust distribution. However, we note that the $1\sigma$ confidence interval nearly encompasses the fixed $\lambda_\mathrm{thick} = 65\,\mu$m assumed previously. We infer a dust mass of $\log_{10}(M_\mathrm{dust}/\mathrm{M}_\odot) = 7.4_{-0.3}^{+0.5}$, which is $\sim6\times$ lower than, albeit still formally consistent with, the value inferred from a fully optically thin treatment. We discuss how our interpretation of REBELS-25 is affected by whether the dust is optically thick in the following section.

\section{Discussion}
\label{sec:discussion}

\subsection{Optically Thickness of REBELS-25}
\label{sec:discussion_optical_depth}
In Section \ref{sec:results}, we presented three MBB fits to the dust SED of REBELS-25, using either fully optically thin dust, marginally optically thick dust with $\lambda_\mathrm{thick} = 65\,\mu$m, or a fit where $\lambda_\mathrm{thick}$ is allowed to vary, resulting in $\lambda_\mathrm{thick} \sim 150\,\mu$m, albeit with large uncertainties. Conclusively distinguishing between these possibilities requires sensitive observations on the blue side of the peak of the dust SED of REBELS-25. For reference, the current Band 9 observations of REBELS-25 have an on-source time of just $\sim25\,$minutes, so deeper observations amounting to a few hours on-source would be highly useful. However, given the expensive nature of high-frequency ALMA observations, a statistical grasp of optical depth effects in high-redshift galaxies is likely out of reach until next-generation facilities such as PRIMA \citep{moullet2023}.

Nevertheless, we can rule out the dust being optically thick at wavelengths much larger than $\lambda_\mathrm{thick} \gtrsim 90\,\mu$m with some confidence -- at least on galaxy-integrated scales -- based on the effect of dust self-absorption on line emission. REBELS-25 was previously detected in \oiiil{} emission \citep{algera2024}, with follow-up high-resolution observations demonstrating the \oiii{} and dust emission to be co-spatial (H.\ Algera et al.\ in prep). Assuming the gas and dust are well-mixed, a high optical depth at the rest-frame wavelength of \oiii{} would suppress the line emission. In particular, the $\lambda_\mathrm{thick} \sim 150\,\mu$m suggested by the fit would imply an optical depth of $\tau(88\,\mu\mathrm{m}) \approx 4$, suppressing the \oiii{} luminosity by a factor of $e^{\tau} \gtrsim50\times$, which is not realistic. As such, the true value of the transition wavelength is likely $\lambda_\mathrm{thick} \lesssim 90\,\mu$m.

However, we can not definitively rule out the presence of dense, optically thick clumps within REBELS-25. \citet{ferrara2022} previously suggested REBELS-25 to be characterized by a two-phase ISM, whereby the dust and UV emission emanate from different physical regions. This is consistent with the spatial offsets observed between the bulk of the dust and UV emission of REBELS-25 by \citet{hygate2023}. While a two-phase, clumpy ISM does not necessarily imply the dust is optically thick, regions that are very opaque to UV photons could -- if dense enough -- also lead to dust self-absorption, on scales far below the current resolution.

In light of these prevailing unknowns, as well as for consistency with the general high-redshift literature, we proceed with our optically thin model as the fiducial one. However, throughout the following discussion, we highlight how our interpretation changes depending on the assumed opacity.

\subsection{Dust Production in REBELS-25}
\label{sec:discussion_dust_production}

Assuming optically thin dust, we measure a high dust mass in REBELS-25 of $\log_{10}(M_\mathrm{dust}/\mathrm{M}_\odot) = 8.2_{-0.4}^{+0.6}$. Based on its stellar mass of $\log_{10}(M_\star/\mathrm{M}_\odot) = 10.3_{-0.2}^{+0.1}$, which was obtained from SED-fitting with a non-parametric star-formation history (\citealt{topping2022}), we infer a high dust-to-stellar mass ratio of $M_\mathrm{dust} / M_\star = 0.8_{-0.5}^{+2.0} \times 10^{-2}$. We note that this is likely to be a lower limit: Stefanon et al.\ (in prep) find a slightly lower (albeit consistent) stellar mass for REBELS-25 of $\log_{10}(M_\star / \mathrm{M}_\odot) = 9.9 \pm 0.2$ assuming a constant star-formation history. If we were to adopt their measurement instead, we would infer an even higher dust-to-stellar mass ratio of  $M_\mathrm{dust} / M_\star = 1.9_{-1.2}^{+6.0} \times 10^{-2}$. In light of this, we adopt the higher stellar mass measurement by \citet{topping2022} as fiducial, emphasizing that a lower mass would require even more rapid dust production.

\subsubsection{Supernova Dust Production}

We start by assuming the dust in REBELS-25 is produced solely through CC-SNe, following the framework introduced in \citet{michalowski2015}. This framework assumes a \citet{chabrier2003} IMF and supposes all stars with masses between $8 \leq M / M_\odot \leq 40$ end up producing dust after exploding in a CC-SN. Under these assumptions, the dust yield per SN $y$ is calculated as $y = 84 \times (M_\mathrm{dust} / M_\star) \approx 0.7_{-0.4}^{+2.3} \mathrm{M}_\odot/\mathrm{SN}$.

Observations of SNe in the local Universe have demonstrated that significant amounts of dust can form in their ejecta ($\sim0.2 - 1.0\,\mathrm{M}_\odot$; \citealt{indebetouw2014,delooze2017,niculescu-duvaz2022}), in agreement with theoretical models (e.g., \citealt{todini2001,marassi2019}). However, it remains unknown how much of this dust is eventually able to enrich the interstellar medium, as a significant fraction (potentially up to $\sim90\,\%$) will subsequently get destroyed by the reverse shock (e.g., \citealt{bianchi2007,kirchschlager2019,kirchschlager2024}; see also the reviews by \citealt{micelotta2018,schneider2023}). As such, the empirically inferred dust yield for REBELS-25 of $y\sim0.7\,\mathrm{M}_\odot$ would require a combination of highly efficient dust production through SNe, and inefficient destruction by the reverse shock.

This need for high dust production efficiencies by SNe can be alleviated if 1) the dust mass is overestimated due to optical depth effects; 2) there are a larger number of SNe per unit stellar mass (i.e., a top-heavy IMF); or 3) there are additional dust production mechanisms beyond SNe.

We first consider the scenario wherein the dust mass is overestimated. If we adopt the fit whereby the dust is optically thick up to a (fixed) wavelength of $\lambda = 65\,\mu$m, we infer a lower dust mass of $M_\mathrm{dust} \approx 10^{8.0}\,\mathrm{M}_\odot$ (Table \ref{tab:fluxes}), resulting in a lower supernova yield of $y = 0.5_{-0.3}^{+1.3}\,\mathrm{M}_\odot$. However, as discussed above, this still requires rather efficient SNe, or a low dust destruction efficiency. Adopting the fit with a variable transition wavelength, which resulted in $\lambda_\mathrm{thick} \approx 150\,\mu$m (Section \ref{sec:results_vary_thick}), we infer a relatively low yield of $y = 0.15_{-0.08}^{+0.39}\,\mathrm{M}_\odot$. As such, we would need the dust to be optically thick out to large wavelengths to recover a yield that is typical of local SNe ($y \sim 0.2\,\mathrm{M}_\odot$; \citealt{schneider2023}), which we argued in Section \ref{sec:discussion_optical_depth} is unlikely.

An alternative way of producing more dust via supernovae without increasing the yield of any individual SN is through a top-heavy initial mass function. The existence of a top-heavy IMF in high-redshift galaxies has long been posited (e.g., \citealt{baugh2005}), and has seen further resurgence in light of recent \textit{JWST} results suggesting an over-abundance of UV-luminous galaxies at $z\gtrsim10$ (e.g., \citealt{harikane2023,trinca2024}). Adopting the top-heavy IMF in \citet{michalowski2015} with a slope of $\alpha = 1.5$ (c.f., a slope of $\alpha = 2.35$ for a \citealt{salpeter1955} IMF), the inferred SNe dust yield for REBELS-25 reduces to $y = 0.5_{-0.3}^{+1.5}\,\mathrm{M}_\odot/\mathrm{SN}$ (assuming fully optically thin dust). However, this yield only modestly reduces the need for efficient dust production in SNe compared to a fiducial Chabrier IMF.

An even more top-heavy IMF, potentially in combination with moderately optically thick dust, could further reduce this yield and thus alleviate tension with observations of local SNe. However, we argue below that a third scenario -- dust production via grain growth in the ISM -- constitutes a more appealing solution in explaining the large dust mass observed in REBELS-25.

\subsubsection{Grain Growth in the ISM}

Grain growth in the interstellar medium consists of two distinct processes: accretion of gas-phase metals onto dust grains, and coagulation of individual dust particles. The former increases the overall dust mass of the system, while at the same time reducing the gas-phase metallicity and altering the size distribution of dust grains (e.g., \citealt{hirashita2012}). Coagulation, on the other hand, alters the grain size distribution while leaving the total dust mass unchanged.

A wide range of studies have demonstrated that the efficiency of grain growth through accretion is strongly dependent on the metallicity \citep{asano2013,devis2019,li2019,galliano2021}. At low metallicities, the lack of available gas-phase metals inhibits efficient grain growth through accretion. However, above a critical metallicity threshold of $Z\gtrsim0.2\,Z_\odot$, rapid grain growth through metal accretion can take place, until a maximum dust-to-metal ratio (DtM) of $\mathrm{DtM} \approx 0.6$ is reached, above which grain growth instead becomes limited by a lack of available metals \citep{li2019,konstantopoulou2022,palla2024}. At low and intermediate redshifts, dust growth via accretion is thought to be the dominant process of dust build-up in galaxies \citep{devis2019,galliano2021,konstantopoulou2023}. In addition, the onset of efficient grain growth is the likely explanation for the observed scaling between the gas-to-dust ratio ($\delta_\mathrm{GDR}$) and metallicity of local galaxies, whereby more metal-rich systems tend to have lower gas-to-dust ratios (e.g., \citealt{remyruyer2014,devis2019}).

\textit{JWST}/NIRSpec Integral Field Unit observations of REBELS-25 suggest it has a high gas-phase metallicity for a $z > 7$ galaxy of $Z \sim 0.9\,Z_\odot$ (L.\ Rowland et al.\ in prep and M.\ Stefanon et al.\ in prep). This supports the scenario where grain growth through metal accretion is efficient in building up its total dust mass. This is corroborated by the recent simulations from \citet{palla2024}, which also suggest grain growth in the ISM to be an important pathway of dust build-up in REBELS galaxies, including in REBELS-25 specifically. Furthermore, adopting the \cii{}-based molecular gas mass of REBELS-25 of $M_\mathrm{gas} = 5.1_{-2.6}^{+5.1} \times 10^{10}\,M_\odot$ (\citealt{hygate2023}; see also \citealt{aravena2024}), we infer a gas-to-dust ratio of $\delta_\mathrm{GDR} = 360_{-280}^{+840}$. Assuming the local scaling relation by \citet{devis2019}, this points towards a similarly high metallicity of $Z > 0.20\,Z_\odot$. While a detailed investigation of the combined dust and metal contents of REBELS-25 is beyond the scope of this paper, the currently available data point towards it being a metal-rich system in which dust growth through accretion is likely to be an efficient process.

In addition to being driven by a high metallicity, the timescale of grain growth is furthermore expected to depend on the ISM density (e.g., \citealt{asano2013}), as a denser ISM facilitates depletion of metals onto dust.
Making use of the recent detection of its CO(7-6) line, A.\ Hygate et al.\ (in prep) perform photo-dissociation region (PDR) modelling to demonstrate REBELS-25 is characterized by a high ISM density of $n(H) \sim 10^{4-5}\,\mathrm{cm}^{-3}$ (see also Y. Fudamoto et al.\ in prep who obtain a similar measurement using far-IR fine structure lines). This is consistent with the simulations by \citet{behrens2018}, which show that high-redshift galaxies contain dense clumps ($n\sim10^4\,\mathrm{cm}^{-3}$) from which most of the infrared emission emanates. If REBELS-25 indeed hosts such a dense, clumpy ISM, conditions are likely favourable for rapid grain growth to take place.

\begin{figure*}
    \centering
     \includegraphics[width=1.0\textwidth]{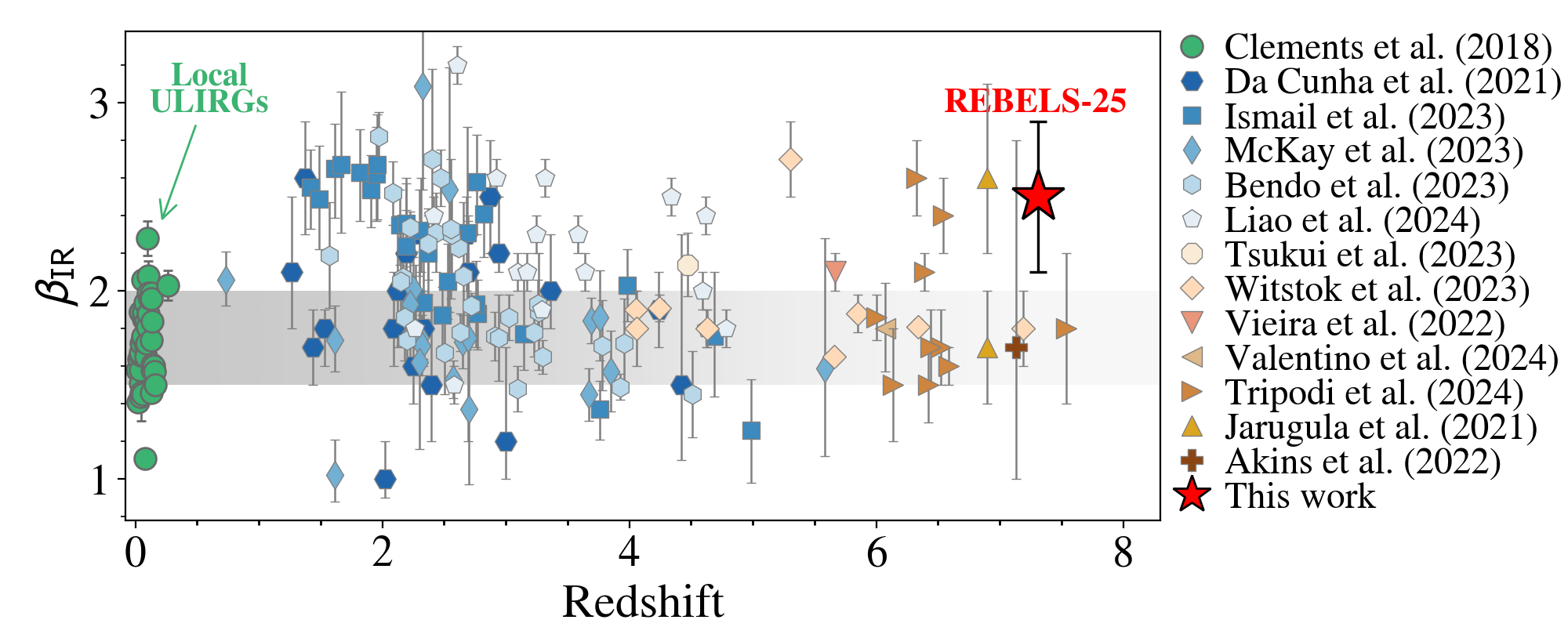}
     \caption{The dust emissivity index ($\beta_\mathrm{IR}$) as a function of redshift. REBELS-25, with $\beta_\mathrm{IR} = 2.5 \pm 0.4$, is indicated via the red star. We further compare to local ULIRGs \citep{clements2018}, dusty star-forming galaxies \citep{dacunha2021,bendo2023,ismail2023,mckay2023,witstok2023b,liao2024}, high-redshift quasars \citep{tsukui2023,witstok2023b,tripodi2024}, the bright two-component system SPT0311-58 at $z=6.90$ \citep{strandet2017,marrone2018,jarugula2021}, and additional individual high-redshift galaxies \citep{vieira2022,akins2022,valentino2024}. The canonical range of $\beta_\mathrm{IR} = 1.5 - 2.0$ is highlighted through the grey shading. Approximately $\sim15\%$ of DSFGs and other high-redshift galaxies and quasars appear to have steep dust emissivity indices ($\beta_\mathrm{IR} \gtrsim 2.5$).}
     \label{fig:beta_redshift}
\end{figure*}

A dense ISM could simultaneously explain the relatively cold dust temperature of REBELS-25. In the presence of a large dust reservoir -- such as in the case of REBELS-25 -- the energy input per unit dust mass is low because of shielding of interstellar radiation, and hence reduces the overall dust heating rate (e.g., \citealt{sommovigo2020}). Lower dust temperatures in shielded environments have also been observed in Galactic objects (e.g., \citealt{hocuk2017}). In such environments, dust grain coagulation is also expected to occur, in addition to accretion \citep{hirashita2012}, resulting in the formation of grains with complex shapes. The far-infrared dust emissivity of these so-called fractal aggregates is thought to increase \citep{stepnik2003,paradis2009}, resulting in a lower equilibrium temperature (e.g., \citealt{kohler2012}). We investigate this possible relation between grain coagulation and the observed dust emissivity index $\beta_\mathrm{IR}$ in Section \ref{sec:discussion_steep_beta}.

\subsection{Dust Emissivity Indices across Cosmic Time}

Our fiducial model of the dust SED of REBELS-25 reveals a dust emissivity index of $\beta_\mathrm{IR} = 2.5 \pm 0.4$. We compare this value to various studies measuring $\beta_\mathrm{IR}$ for galaxy samples at lower redshift in Figure \ref{fig:beta_redshift}, limiting ourselves to sources with spectroscopic redshifts. We adopt measurements of local ULIRGs from \citet{clements2018}, dusty star-forming galaxies (DSFGs) at intermediate redshift from a variety of recent studies \citep{dacunha2021,bendo2023,ismail2023,mckay2023,liao2024} as well as $z\gtrsim6$ quasars from \citet{tripodi2024}. Moreover, we compare to several additional sources -- both galaxies and quasars -- at $4 \lesssim z \lesssim 7$ for which dust emissivity indices have recently been measured \citep{akins2022,vieira2022,tsukui2023,witstok2023b,valentino2024}. \\

The value of $\beta_\mathrm{IR}$ obtained for REBELS-25 is steeper than what is generally measured in samples of local galaxies, for which an average value of $\langle \beta_\mathrm{IR} \rangle \approx 1.7 - 1.8$ has been determined \citep{cortese2014,clements2018}. Higher redshift constraints on $\beta_\mathrm{IR}$ are mostly from DSFGs, which show a variety of emissivity indices ranging from an average $\langle \beta_\mathrm{IR} \rangle \approx 1.8 - 2.4$ with significant scatter \citep{casey2021,dacunha2021,cooper2022,bendo2023,ismail2023,mckay2023,liao2024}. While values of $\beta_\mathrm{IR} \gtrsim 2.5$ do not appear to be typical for DSFGs, they are also not uncommon, with $\sim15\%$ of DSFGs in Figure \ref{fig:beta_redshift} showing a similarly steep or steeper dust emissivity index than REBELS-25.

However, we note that comparing fitted dust emissivity indices among different studies is complicated by the variety of fitting techniques utilized in the literature. For example, \citet{mckay2023} show that adopting optically thick dust can steepen the measured $\beta_\mathrm{IR}$, which may partially explain the relatively steep dust emissivity indices ($\langle \beta_\mathrm{IR} \rangle > 2$) observed in DSFG studies adopting $\lambda_\mathrm{thick} = 200\,\mu$m \citep{casey2021,cooper2022}. Indeed, in our MBB fit where the transition wavelength is left as a free parameter, we infer $\lambda_\mathrm{thick} \approx 150\,\mu\mathrm{m}$ and a steeper $\beta_\mathrm{IR} \approx 2.7$ (Table \ref{tab:fluxes}). This highlights that the steep dust emissivity index we infer for REBELS-25 is robust against assumptions regarding the optical depth. 

Steep values of $\beta_\mathrm{IR}$ are furthermore robust against the presence of multi-temperature dust components within the galaxy. While our fitting formalism approximates REBELS-25 as a single-temperature modified blackbody, a realistic galaxy will undoubtedly be characterized by a distribution of temperatures -- warm(er) in star-forming regions and cold(er) in the diffuse ISM \citep{behrens2018,liang2019,sommovigo2020}. However, the effect of such multi-temperature dust is instead to flatten the observed emissivity index, rather than steepen it (e.g., \citealt{hunt2015,lamperti2019}). The cold dust temperature recovered for REBELS-25 may further suggest the effects of such temperature mixing to be limited, as any warm dust components that may be present do not dominate the emission at the wavelengths probed in this work. This would moreover imply that the current observations are already probing the bulk of the total dust mass in REBELS-25. \\ 

Beyond $z\gtrsim5$, dust emissivity index measurements become scarce, due to the apparent dearth of infrared-bright star-forming galaxies at this epoch. \citet{akins2022} infer a value of $\beta_\mathrm{IR} = 1.7_{-0.7}^{+1.1}$ for the $z=7.13$ galaxy zD1, although the dust continuum in this source is not detected redwards of ALMA Band 6 (rest-frame $\lambda_\mathrm{rest} \approx 160\,\mu\mathrm{m}$), resulting in only loose constraints on the emissivity index. The two components of the uniquely massive system SPT0311-58 at $z=6.900$, on the other hand, are detected in dust continuum emission in ALMA Bands 3 and 4, enabling a dust emissivity index measurement \citep{strandet2017,marrone2018,jarugula2021}. Re-fitting the ALMA photometry presented in \citet{jarugula2021} with a MBB that is optically thick out to $\lambda_\mathrm{thick} = 100\,\mu$m -- motivated by the large dust mass of the two components -- we measure a dust emissivity of $\beta_\mathrm{IR,W} = 1.7 \pm 0.3$ and $\beta_\mathrm{IR,E} = 2.6_{-0.4}^{+0.5}$ for the bright western and fainter eastern components, respectively. This is consistent with the typical values measured for DSFGs at lower redshift \citep{dacunha2021}, of which this massive dusty system is likely a particularly early example. Additional measurements of $\beta_\mathrm{IR}$ are available for distant, far-infrared luminous quasars, with \citet{tripodi2024} recently having compiled the dust SEDs of 10 quasars at $z\gtrsim6$. Among this sample, they find dust emissivity indices spanning the range $\beta_\mathrm{IR} = 1.5 - 2.6$, with three quasars having $\beta_\mathrm{IR} > 2$. Moreover, \citet{witstok2023b} recently measured the dust emissivity indices for a compilation of $4 \lesssim z \lesssim 7.5$ galaxies and quasars, for which they find a typical $\beta_\mathrm{IR} = 1.8 \pm 0.3$ and only a single outlier with $\beta_\mathrm{IR} > 2.5$. However, we note that their work assumes a relatively narrow Gaussian prior on the dust emissivity index with a mean and standard deviation of $1.8$ and $0.25$, respectively (see also \citealt{valentino2024}), which is expected to lead to tighter constraints compared to a uniform prior like the one adopted in this work.

The above studies suggest that, while not the norm, values of $\beta_\mathrm{IR} \gtrsim 2$ are not particularly uncommon. At present, however, studies of dust properties at $z\gtrsim6$ remain limited to small numbers of the brightest objects. A detailed comparative study of the dust properties of reionization-era galaxies therefore remains out of reach, and larger galaxy samples with well-measured dust emissivity indices at early cosmic epochs are required to assess whether the average dust properties of galaxies may change across cosmic time. In the following section, we therefore focus on the steep dust emissivity index in REBELS-25 specifically, and discuss a possible link to grain growth taking place in its interstellar medium.

\subsection{The Dust Mass and Emissivity Index of REBELS-25}
\label{sec:discussion_steep_beta}

Observations of the Milky Way show that the submillimeter dust emissivity index steepens in cold, dense regions \citep{stepnik2003,paradis2009}. The preferred explanation for this finding is the altered emission properties of coagulated dust in a dense ISM. In particular, theoretical modeling by \citet{kohler2015} suggests that both accretion of metals and coagulation of dust grains can steepen the dust emissivity index, with coagulation being required to reach steep values of $\beta_\mathrm{IR} > 2$. While coagulation itself does not lead to an increase in the overall dust mass, both coagulation and accretion occur preferentially in the dense ISM, and are thus likely to take place in tandem \citep{hirashita2012}. This explanation of a steeper $\beta_\mathrm{IR}$ being the result of both grain growth processes is appealing for REBELS-25, given that grain growth through metal accretion is likely necessary to explain its large dust mass. At the same time, however, \citet{kohler2015} find that the overall emissivities of these coagulated grains increase by a factor of a few, especially upon the formation of icy mantles on the coagulated grain surfaces. If indeed the case, such coagulated grains will thus be more luminous than standard dust grains given a fixed total dust mass, and may hence cause the observationally inferred dust mass to be overestimated.

\citet{ysard2019} show that a flattening dust emissivity index at millimeter wavelengths -- beyond the wavelength range probed in this work -- may also be a signature of large dust grains. Such a flattening has indeed been observed at long wavelengths in compact, dusty cores within the Galaxy \citep{juvela2015}, although no clear flattening is seen on integrated scales in DSFGs \citep{hirashita2023}. At sub-millimeter wavelengths, the models from \citet{ysard2019} instead do not show strong variation in terms of their dust emissivity indices. Indeed, they argue that extracting detailed information on dust grain sizes and composition from (sub)millimeter observations alone is a complicated and inherently uncertain endeavour. 

While it would thus be compelling to argue that the steep dust emissivity index observed in REBELS-25 supports the onset of grain growth in its ISM, the truth of the matter is that the relation between the dust emissivity index and physical properties of dust grains remains nebulous. Instead, to more clearly highlight whether dust properties in the early Universe are indeed different from those in galaxies at lower redshift, the assembly of larger samples of distant galaxies with well-constrained dust emissivity indices is essential. As such, we outline in the next section how to best do this with ALMA.

\begin{figure*}
    \centering
     \includegraphics[width=0.9\textwidth]{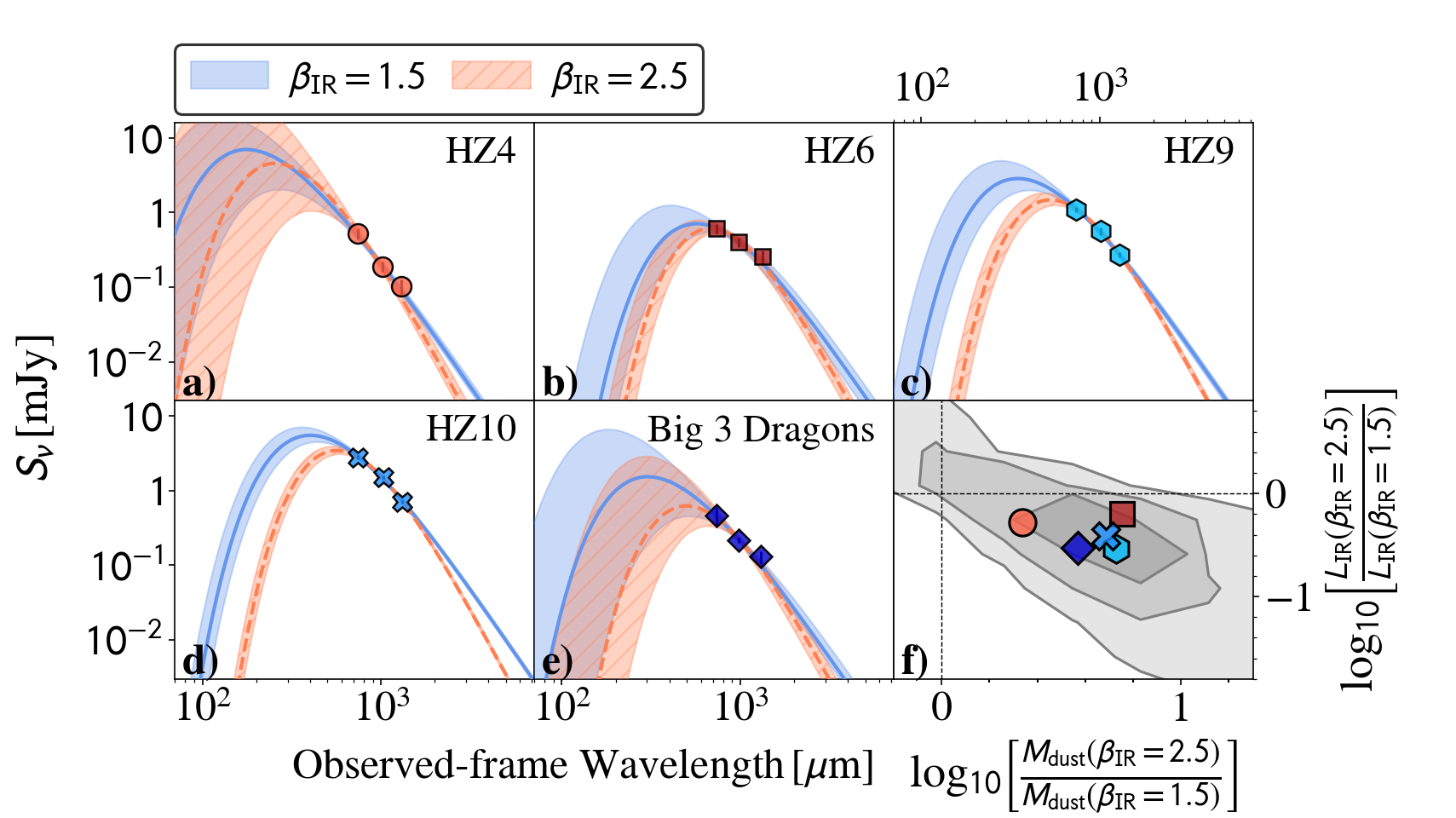}
     \caption{\textbf{Panels a) - e):} Optically thin MBB fits to five galaxies at $z\approx5.5 - 7.2$ with detections in ALMA Bands 6, 7 and 8 \citep{hashimoto2019,faisst2020,sugahara2021}. The markers represent the observed flux densities, and are generally larger than the uncertainties. A fixed emissivity index of $\beta_\mathrm{IR} = 1.5$ (blue) or $\beta_\mathrm{IR} = 2.5$ (orange) is assumed in the MBB fits. While the data can be well described by a fit using either value of $\beta_\mathrm{IR}$, the resulting inferred dust SEDs -- and thus the inferred dust temperatures, masses and IR luminosities -- are visibly different. \textbf{Panel f):} Difference in the inferred IR luminosity and dust mass for the 5 literature sources, given the two assumed emissivity indices. The background shading shows the combined 2D posterior distribution from the MBB fits, with the markers showing the median difference (symbols are the same for the five sources in panels a - e). Assuming a steeper $\beta_\mathrm{IR}=2.5$ results in dust masses that are $\sim0.7\,$dex higher compared to $\beta_\mathrm{IR}=1.5$, while IR luminosities are $\sim0.4\,$dex lower. As such, a lack of knowledge of $\beta_\mathrm{IR}$ results in substantial systematic uncertainties on the recovered dust masses and obscured SFRs.}
     \label{fig:literatureSEDs}
\end{figure*}

\section{Measuring Dust Emissivity Indices at High Redshift}
\label{sec:mockfitting}

\subsection{The Need for Probing the Rayleigh-Jeans Tail}
\label{sec:literaturefit}

Conventional wisdom dictates that three ALMA measurements are required to constrain the three key dust parameters -- dust temperature, mass and emissivity index. However, due to degeneracies between these quantities, the accuracy with which $\beta_\mathrm{IR}$ can be recovered depends strongly on which particular frequencies are observed. As we will demonstrate in Section \ref{sec:discussion_mockfitting}, two ingredients are crucial for an accurate measurement of the emissivity index: 1) a measurement sampling the Rayleigh-Jeans tail of the dust emission, and 2) a measurement sampling the peak of the dust SED. Together, this set provides a long lever arm capable of breaking the pesky degeneracy between $T_\mathrm{dust}$ and $\beta_\mathrm{IR}$.

In the absence of such observations, a value of $\beta_\mathrm{IR}$ must be assumed a priori, leading to potentially significant systematic uncertainties. We showcase this by fitting optically thin MBB models to five $z \gtrsim 5$ galaxies in the literature with continuum detections in ALMA Bands 6, 7 and 8 (Fig.\ \ref{fig:literatureSEDs}); four galaxies from \citet{faisst2020} and the Big Three Dragons \citep{hashimoto2019,sugahara2021}.\footnote{We note that, at the time of finalizing this manuscript, \citet{villanueva2024} presented new ALMA Band 9 measurements of the $z=5.66$ galaxy HZ10, which is part of the sample in \citet{faisst2020}.} Assuming either a fixed $\beta_\mathrm{IR} = 1.5$ or $\beta_\mathrm{IR} = 2.5$ results in a good fit to the data, given the relatively short wavelength range spanned by the three bands. Applying the Bayesian Information Criterion (BIC), the fits are statistically indistinguishable ($|\Delta\mathrm{BIC}| < 2.5$). However, assuming the steeper value of $\beta_\mathrm{IR} = 2.5$, inferred dust temperatures decrease by a systematic $\Delta T_\mathrm{dust} = 12 - 45\,$K, dust masses increase by $\Delta \log(M_\mathrm{dust}/M_\odot) = 0.34 - 0.76\,$dex and infrared luminosities decrease by $\Delta \log(L_\mathrm{IR}/L_\odot) = 0.20 - 0.55\,$dex, compared to an assumed $\beta_\mathrm{IR} = 1.5$. As such, adopting a fixed value for $\beta_\mathrm{IR}$ can lead to significant systematic uncertainties in the inferred dust masses and obscured SFRs of distant galaxies.

\subsection{Constraining Dust SEDs with Mock ALMA Observations}
\label{sec:discussion_mockfitting}

\begin{figure*}
    \centering 
    \includegraphics[width=1.0\textwidth]{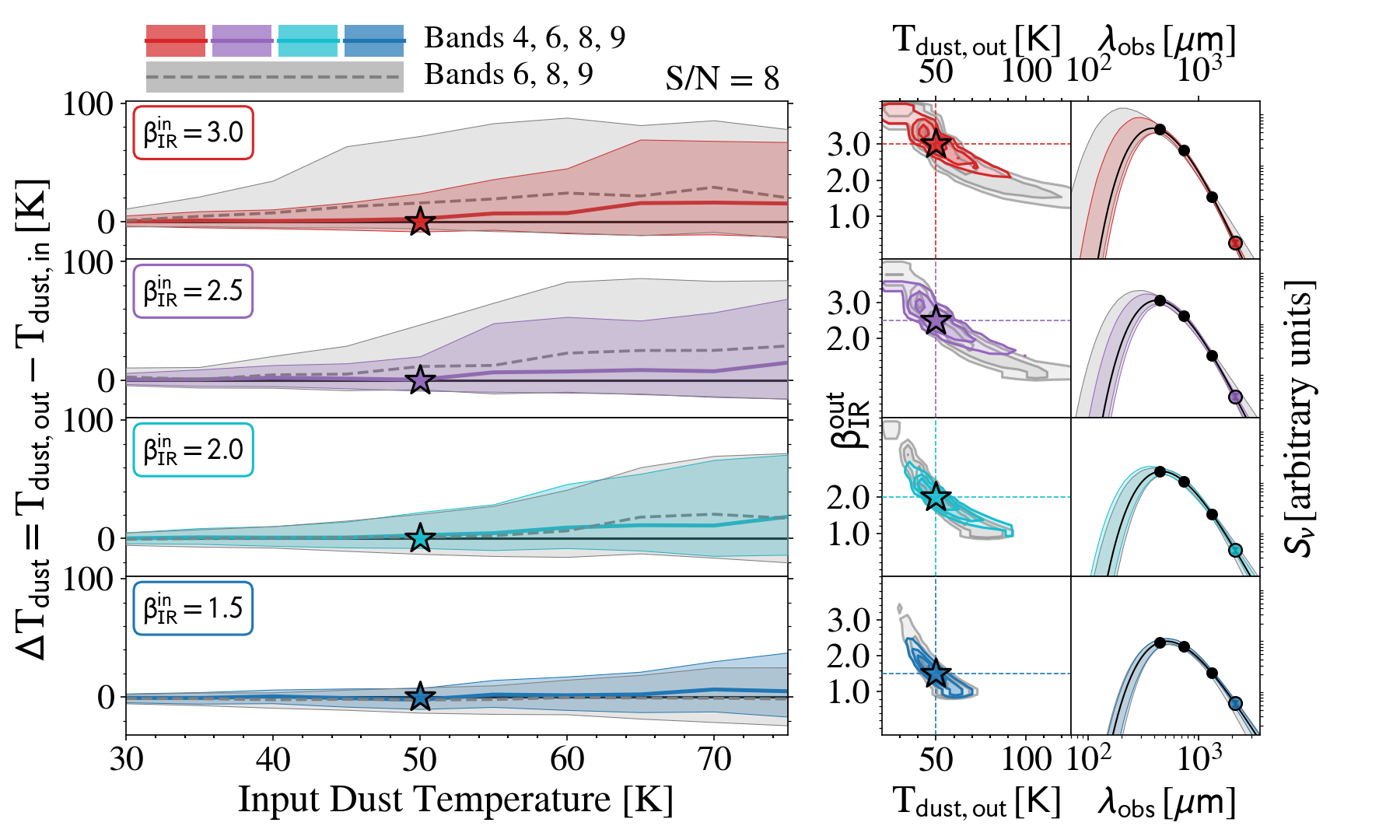}
    \caption{
    Constraints on the dust SED of a simulated $z=7$ galaxy across a range of possible dust temperatures and emissivity indices. The various rows assume a range of input values $1.5 \leq \beta_\mathrm{IR} \leq 3.0$, with the colors showing the constraints for a four-band ALMA setup (Bands 4, 6, 8 and 9) and the greyscale indicating a three-band setup (omitting Band 4). A $\mathrm{S/N} = 8$ is assumed for each band. \textbf{Left column:} Accuracy with which the dust temperature can be constrained, as a function of input dust temperature. The solid colored (dashed grey) line indicates the median value of $\Delta T_\mathrm{dust}$ for the 4-band (3-band) ALMA setup, with the shaded regions indicating the $1\sigma$ confidence intervals. The uncertainties increase for sources with hotter dust and/or a steeper $\beta_\mathrm{IR}$, given that this causes the peak of the SED to shift beyond Band 9. \textbf{Middle column:} Simultaneous constraints on $\beta_\mathrm{IR}$ and $T_\mathrm{dust}$ for an input $T_\mathrm{dust}=50\,$K (marked with a star). The confidence intervals shrink considerably when Band 4 is included in the fit, in particular for steeper values of $\beta_\mathrm{IR}$. \textbf{Right column:} Expected constraints on the dust SED of the simulated galaxy given the two ALMA setups for an input $T_\mathrm{dust}=50\,$K. The constraints at long wavelengths improve significantly when Band 4 is added to the fit, while the constraints at short wavelengths similarly improve when $\beta_\mathrm{IR}$ is steep. }
    \label{fig:mock_fitting}
\end{figure*}

\begin{figure*}
    \centering 
    \includegraphics[width=0.33\textwidth]{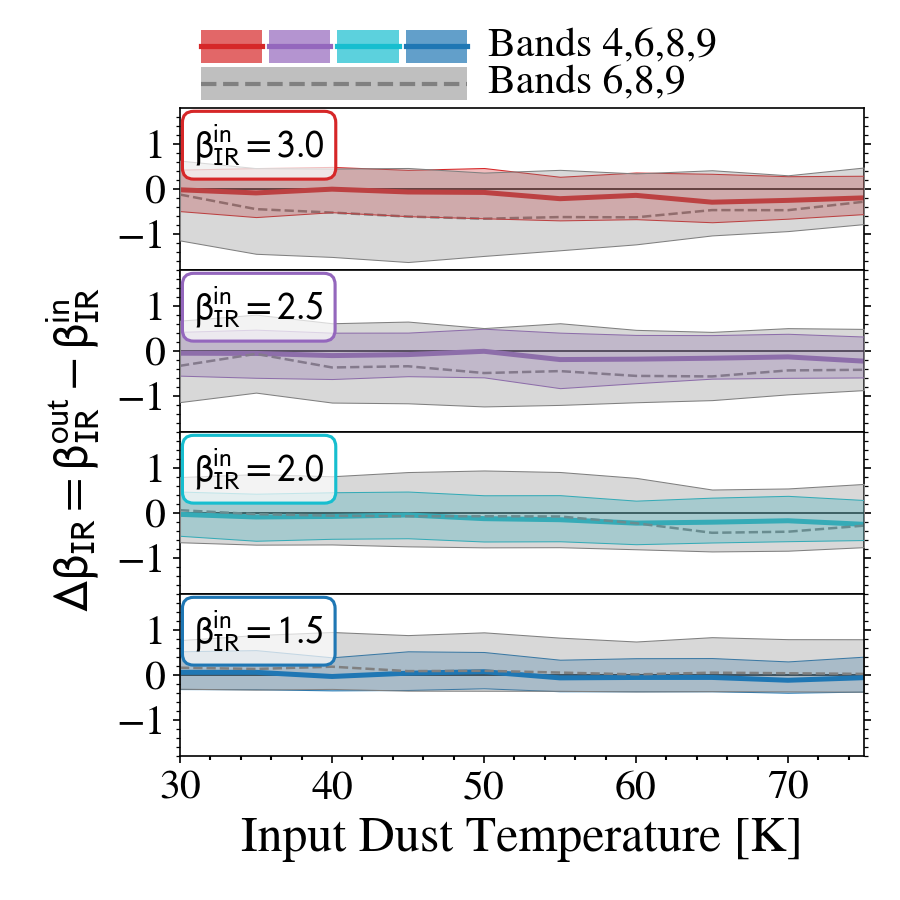}
    \includegraphics[width=0.33\textwidth]{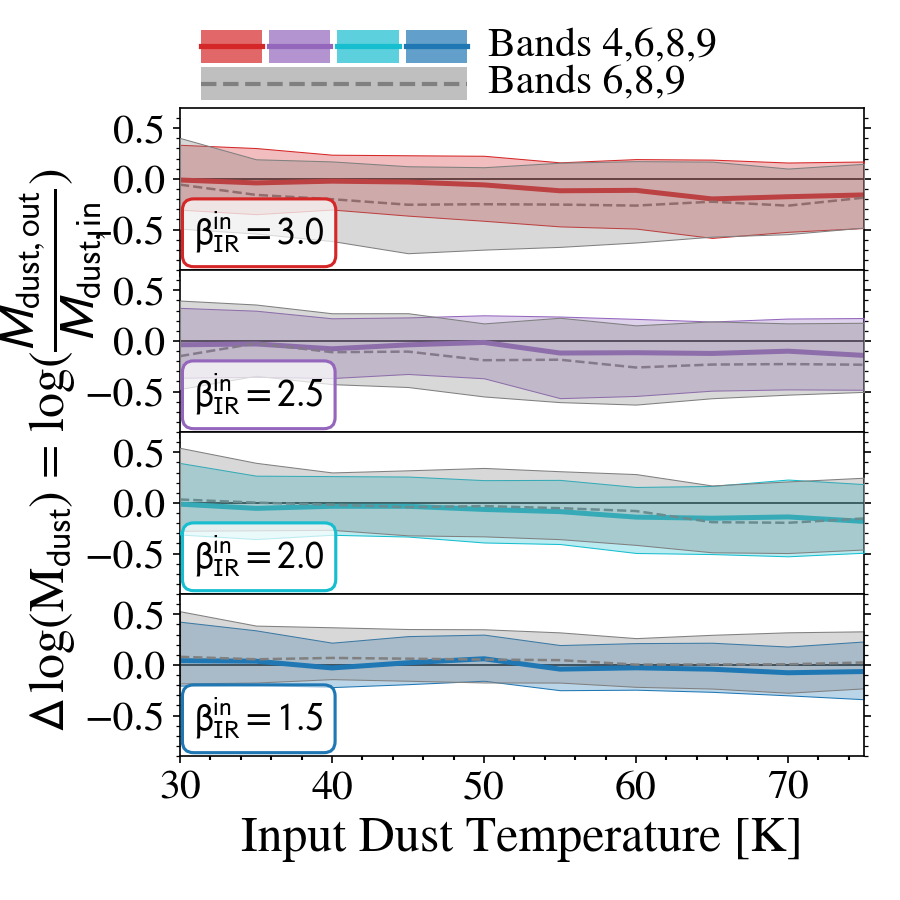}
    \includegraphics[width=0.33\textwidth]{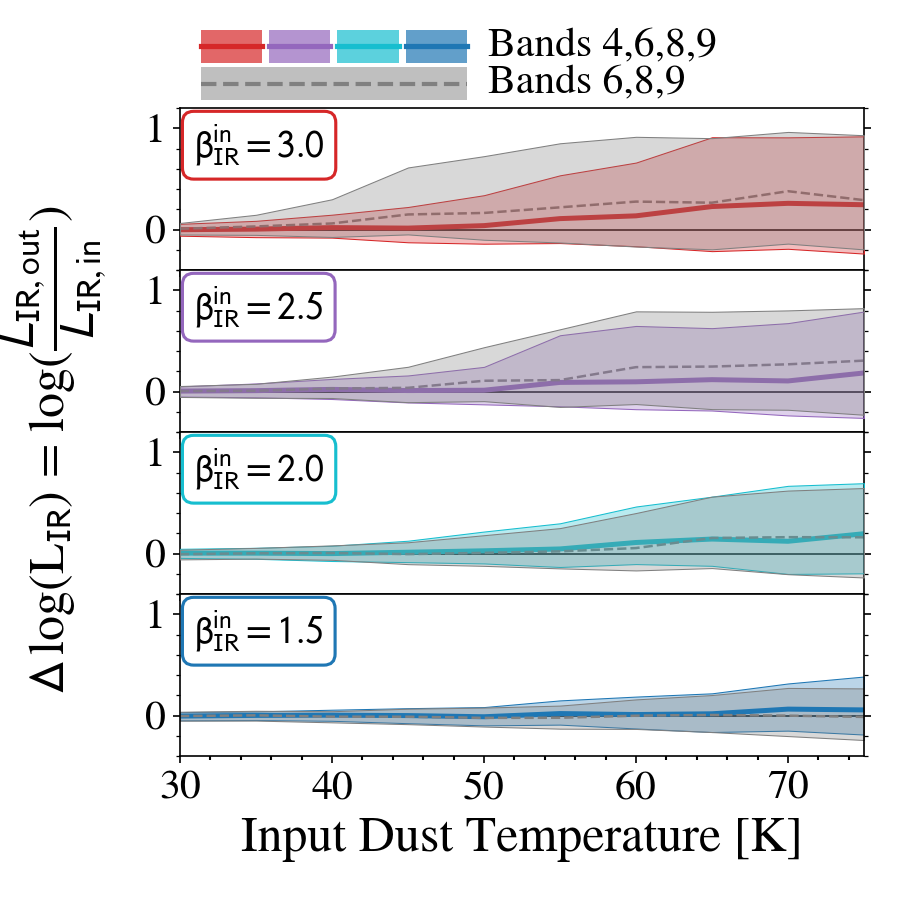}
    \caption{Constraints on the dust emissivity index (left), dust mass (middle) and IR luminosity (right) as a function of the input dust temperature, for a set of MBB fits of a simulated $z=7$ galaxy. The different rows correspond to four different input values of $1.5 \leq \beta_\mathrm{IR} \leq 3.0$. The solid line and colored shading show the constraints using a four-band setup (Bands 4, 6, 8, 9) for an S/N = 8 per band, while the dashed grey line and grey shading show the results for a three-band setup, omitting Band 4. Without coverage of the Rayleigh-Jeans tail, $\beta_\mathrm{IR}$ cannot accurately be constrained, which results in larger uncertainties on the other dust parameters -- especially when the input value of $\beta_\mathrm{IR}$ is steep.}
    \label{fig:mock_fitting_v2}
\end{figure*}

Having established that measuring $\beta_\mathrm{IR}$ is important to accurately constrain the dust SEDs of high-redshift galaxies, we next discuss how to do this reliably with multi-band ALMA observations. In practice, this tends to be a multi-step process, whereby slowly more observing bands are gathered for a galaxy of interest. At $z \approx 6-7.5$, galaxies are most typically targeted at Bands 6 and 8 first to simultaneously observe their \ciil{}, \oiiil{}, and underlying dust continuum emission (e.g., \citealt{harikane2020,witstok2022,mitsuhashi2023a,algera2024}), only after which more bands are added to the mix. As such, we here focus on a hypothetical galaxy at $z=7$ with existing observations in ALMA Bands 6 and 8, and investigate to what extent adding additional bands helps constrain its dust SED through fitting mock ALMA observations. 

The details of this fitting process are given in Appendix \ref{app:mock_comparison}, where we additionally demonstrate that adding Band 9 to this dual-band setup of ALMA Bands 6 and 8 is the optimal choice if one is interested in obtaining tight constraints on galaxy dust masses and IR luminosities (Fig.\ \ref{fig:mockfitting_setup}). In what follows, however, we focus our attention on the Rayleigh-Jeans tail of the dust SED. In Appendix \ref{app:mock_comparison} we find that -- once Band 9 observations are available -- constraining the Rayleigh-Jeans tail via observations in ALMA Band 4 becomes the most efficient observing strategy. As such, we simulate the expected constraints on the dust SED of our hypothetical $z=7$ galaxy, assuming either a three-band setup of ALMA Bands [6, 8, 9], or a four-band setup further including ALMA Band 4. We adopt a fixed S/N = 8 for each of the bands, as we demonstrate in Appendix \ref{app:mock_lowsnr} that a lower S/N = 4 per band is generally not sufficient in order to obtain tight constraints on the dust parameters (Fig.\ \ref{fig:mock_fitting_lowsnr}).

In practice, the continuum S/N of galaxies may vary significantly between individual bands (e.g., \citealt{faisst2020,witstok2022,algera2024}). For example, attaining S/N = 8 is likely highly time-consuming at ALMA Bands 4 and 9, while it is reasonable to exceed this S/N at Bands 6 and 8. As such, the simulations presented below show only a general case. Accurately forecasting the expected constraints on $\beta_\mathrm{IR}$ for any particular galaxy requires considering the S/N of previously taken observations, as well as obtaining a rough estimate of its dust temperature (presumably through a fit with $\beta_\mathrm{IR}$ kept fixed to a fiducial value). In the future, the ALMA Wideband Sensitivity Upgrade \citep{carpenter2022}, currently planned for 2030, will increase the continuum observing speed of ALMA by $\gtrsim3\times$, facilitating more efficient sampling of the dust SEDs of distant galaxies. \\

We show the accuracy with which the dust temperature can be constrained from two MBB fits to mock ALMA photometry in Bands [6, 8, 9] and [4, 6, 8, 9] across a range of input dust temperatures in Figure \ref{fig:mock_fitting}. In addition, we show the recovered dust mass, emissivity index and infrared luminosity in Figure \ref{fig:mock_fitting_v2}. As expected, we find that including Band 4 in the fitting is crucial for accurate constraints on the dust emissivity index, regardless of the input value of $\beta_\mathrm{IR}$ (left panel of Fig.\ \ref{fig:mock_fitting_v2}). Without Band 4, $\beta_\mathrm{IR}$ will be systematically underestimated when the true $\beta_\mathrm{IR} \gtrsim 2$, while for a shallower $\beta_\mathrm{IR}$ the flat prior we adopt ($\beta_\mathrm{IR} \in [1, 4]$) ensures the fitted dust emissivity index is reasonably close to the input value.\footnote{Adopting a wider prior such as $\beta_\mathrm{IR} \in [0, 5]$ without adequate sampling of the Rayleigh-Jeans tail both introduces larger systematic offsets between the input and fitted parameters, and increases the overall parameter uncertainties.} However, irrespective of the true dust emissivity index of the simulated galaxy, uncertainties on $\beta_\mathrm{IR}$ are large without Band 4 observations constraining the Rayleigh-Jeans tail of the dust SED.

Moreover, we find that including Band 4 in the fitting process significantly improves the constraints on the dust temperature, dust mass and IR luminosity for a galaxy with a high $T_\mathrm{dust}$ and/or steep $\beta_\mathrm{IR}$, as these cause the peak of the MBB to shift beyond Band 9. In this case, an accurate measurement of the Rayleigh-Jeans tail can help break the degeneracy between the dust temperature and emissivity index, even when the peak of the dust SED is not adequately probed. 

Focusing on a specific case with an input value of $T_\mathrm{dust} = 50\,$K (middle and right columns of Fig. \ref{fig:mock_fitting}), we find that the uncertainties on $T_\mathrm{dust}$ shrink by a factor of $\gtrsim2\times$ when Band 4 data are included and the dust emissivity index is steep ($\beta_\mathrm{IR} \gtrsim 2.5$). This suggests that the dust SEDs of galaxies with a `canonical' dust emissivity index of $\beta_\mathrm{IR} \approx 1.5 - 2.0$ will not be significantly better constrained when Band 4 is included. However, the catch is that one does not know whether the value of $\beta_\mathrm{IR}$ is indeed canonical until the Rayleigh-Jeans tail of the source of interest is actually observed via observations at Band 4 (or a similar frequency).

Nevertheless, we find that even in the absence of observations probing the Rayleigh-Jeans regime, the systematic uncertainties on the dust mass and infrared luminosity (i.e., the offset between the input and fitted parameters) are relatively small (middle and right panels of Fig.\ \ref{fig:mock_fitting_v2}). As such, a three-band setup comprising ALMA Bands 6, 8 and 9 is likely to be sufficient to constrain the dust mass budget or obscured star formation rate density at high redshift, unless high dust temperatures ($T_\mathrm{dust} \gtrsim 60\,$K) and/or steep dust emissivity indices ($\beta_\mathrm{IR} \gtrsim 2.5$) turn out to be common for distant galaxies.

Finally, we remark that we have purposefully adopted a broad, flat prior on $\beta_\mathrm{IR} \in [1, 4]$ in our fitting framework. In the literature, typically more informative (e.g., Gaussian) priors are adopted when fitting the dust SEDs of high-redshift galaxies (e.g., \citealt{faisst2020,mitsuhashi2023a,algera2024}), in an attempt to marginalize across $\beta_\mathrm{IR}$ when the coverage of the Rayleigh-Jeans tail is insufficient to accurately constrain it. However, because the dust properties of reionization-era galaxies are potentially different from those of galaxies at lower redshifts, and because laboratory experiments indicate some dust analogues may indeed have steep emissivity indices (\citealt{inoue2020,ysard2024}, and references therein), we here follow a conservative approach by adopting a wide, flat prior. While adopting a more informative prior on $\beta_\mathrm{IR}$ is useful in practice when limited to observations with relatively poor S/N, it will have the side effect of limiting one's discovery space to dust emissivity indices more in line with the canonical $\beta_\mathrm{IR} \sim 1.5 - 2$. This may result in more exotic values of $\beta_\mathrm{IR}$ -- which may provide particularly useful insights into the properties of dust in the early Universe -- being overlooked. As such, a more informative prior should not be regarded as an adequate substitute for high-S/N, multi-band observations.

\subsection{Dust-based Gas Mass Measurements with ALMA and JWST}
At low and intermediate redshifts, several studies have advocated for the use of dust mass as a tracer of the total molecular gas mass in massive galaxies (e.g., \citealt{magdis2012,scoville2016,dunne2022}). This is particularly useful in light of the most commonly adopted tracer of molecular gas -- the CO(1-0) line -- being difficult to observe in the early Universe due to its intrinsic faintness and low contrast against, as well as heating by, the CMB (e.g., \citealt{friascastillo2023}). In addition, higher-J CO transitions have also proven difficult to observe at $z\gtrsim 6$ (e.g., \citealt{ono2022,hashimoto2023}), resulting in a shift in the literature towards using the significantly more luminous \ciil{} line to infer the molecular gas contents of distant galaxies (e.g., \citealt{zanella2018,dessauges-zavadsky2020,madden2020,aravena2024}). However, the interpretation of \cii{} as a gas mass tracer remains uncertain in the early Universe, due to a lack of direct calibrations against CO lines, and its simultaneously observed tight scaling relation with star formation rate (e.g., \citealt{delooze2014,schaerer2020}). 

Dust emission, therefore, can be used as an alternative tracer of molecular gas in the early Universe. Provided the dust mass is accurately constrained through multi-band ALMA observations, the primary uncertainty in its application is the gas-to-dust ratio, which from local galaxies is known to be a strong function of metallicity (e.g., \citealt{remyruyer2014,devis2019}). However, metallicity measurements at high-redshift are now readily obtained with the \textit{JWST} (e.g., \citealt{nakajima2023}), enabling a powerful synergy between ALMA and the \textit{JWST} in measuring the molecular gas contents of distant galaxies, once accurate constraints on the far-IR dust properties of sufficiently large galaxy samples are available (see also \citealt{palla2024}).

\section{Conclusions}
\label{sec:conclusions}

We present new simultaneous constraints on the dust mass, temperature, and emissivity index of REBELS-25 -- a massive, UV-luminous star-forming galaxy at $z = 7.31$ -- through six-band ALMA continuum observations. We augment existing detections of the galaxy in Bands 6 and 8 with newly taken observations in Bands 3, 4, 5 and 9. REBELS-25 is detected at $4.3 - 16.3\sigma$ in all bands except for Band 9, where the resulting upper limit helps to constrain its dust SED.

We fit an optically thin modified blackbody to the ALMA continuum measurements of REBELS-25, and find a low dust temperature of $T_\mathrm{dust} = 32_{-6}^{+9}\,$K, in agreement with the $T_\mathrm{dust}$ previously inferred by \citet{algera2024} based on two ALMA measurements. We further measure a high dust mass of $\log_{10}(M_\mathrm{dust}/\mathrm{M}_\odot) = 8.2_{-0.4}^{+0.6}$ for REBELS-25, as well as a steep dust emissivity index of $\beta_\mathrm{IR} = 2.5 \pm 0.4$. If produced solely through supernovae, this dust mass implies a high yield of $y=0.7_{-0.4}^{+2.3}\,M_\odot/\mathrm{SN}$. While it is possible to slightly reduce this yield by invoking optically thick dust, our analysis suggests the SED is unlikely to be sufficiently optically thick to significantly alter the dust mass measurement. Invoking a top-heavy IMF would also slightly reduce the required supernova yields, however, additional dust production through means other than SNe provides a more appealing solution. We argue that both the high dust mass and steep $\beta_\mathrm{IR}$ inferred for REBELS-25 may be indicative of grain growth through metal accretion and coagulation taking place in its ISM, which is likely particularly efficient given its dense and metal-rich nature. As such, grain growth may constitute a key dust production mechanism in the early Universe, in particular in massive, metal-rich and dense galaxies similar to REBELS-25.

Our analysis moreover suggests that measuring dust emissivity indices in high-redshift galaxies is essential to gain insight into their dust properties, and hence into early dust production pathways. While the direct relation between $\beta_\mathrm{IR}$ and dust properties is far from trivial, obtaining larger samples of distant galaxies with well-sampled dust SEDs is required to assess whether the average dust properties of galaxies evolve towards the earliest cosmic epochs. Furthermore, we demonstrate how accurately measuring $\beta_\mathrm{IR}$ can help mitigate systematic uncertainties in the inferred dust temperatures, masses and infrared luminosities of distant galaxies. We show that such measurements require combined observations of the peak and Rayleigh-Jeans tail of the infrared SED at high redshift. While the optimal ALMA setup should be considered on a per-source basis, using previously obtained flux measurements as a prior, for the $z\sim7$ galaxy population a combination of observations in ALMA Bands 4 and 9 constitutes an efficient means to constrain dust SEDs in the early Universe. Once such accurate constraints are in place, dust continuum observations from ALMA combined with metallicity measurements from \textit{JWST} will provide a robust way of inferring the molecular gas masses of distant galaxies.

\section*{Acknowledgements}
We thank the referee for their useful comments and suggestions that significantly improved the clarity of this work. This paper makes use of the following ALMA data: \\ ADS/JAO.ALMA\#2017.1.01217.S, ADS/JAO.ALMA\$2019.1.01634.L, ADS/JAO.ALMA\#2021.1.00318.S, ADS/JAO.ALMA\#2021.1.01495.S, ADS/JAO.ALMA\#2022.1.01324.S, ADS/JAO.ALMA\#2022.1.01384.S. 

ALMA is a partnership of ESO (representing its member states), NSF (USA) and NINS (Japan), together with NRC (Canada), MOST and ASIAA (Taiwan), and KASI (Republic of Korea), in cooperation with the Republic of Chile. The Joint ALMA Observatory is operated by ESO, AUI/NRAO and NAOJ.

This work was supported by NAOJ ALMA Scientific Research Grant Code 2021-19A (H.S.B.A. and H.I.). E.d.C. gratefully acknowledges support from the Australian Research Council: Future Fellowship FT150100079, Discovery Project DP240100589, and the ARC Centre of Excellence for All Sky Astrophysics in 3 Dimensions (ASTRO 3D; project CE170100013). J.H. acknowledges support from the ERC Consolidator Grant 101088676 (VOYAJ) and the VIDI research programme with project number 639.042.611, which is (partly) financed by the Netherlands Organisation for Scientific Research (NWO).

\section*{Data Availability}
All ALMA data used in this work are available via the ALMA science archive (\url{https://almascience.nrao.edu/aq/}). The data underlying this article will be made available upon reasonable request to the corresponding author.



\bibliographystyle{mnras}
\bibliography{main} 


\appendix

\section{Optically Thick MBB Fitting}
\label{app:thick}

\begin{figure*}
    \centering
    \includegraphics[width=0.49\textwidth]{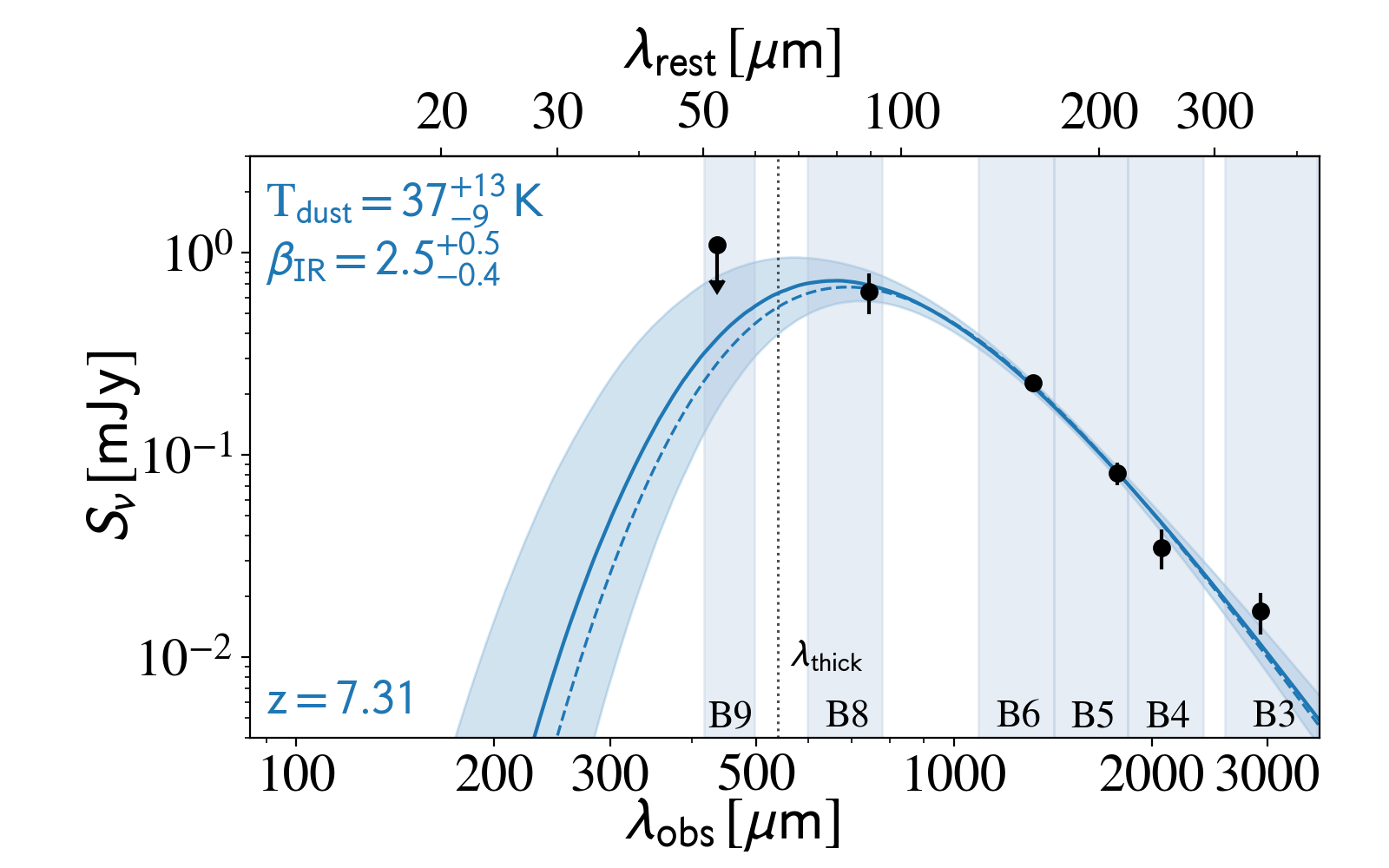}
    \includegraphics[width=0.49\textwidth]{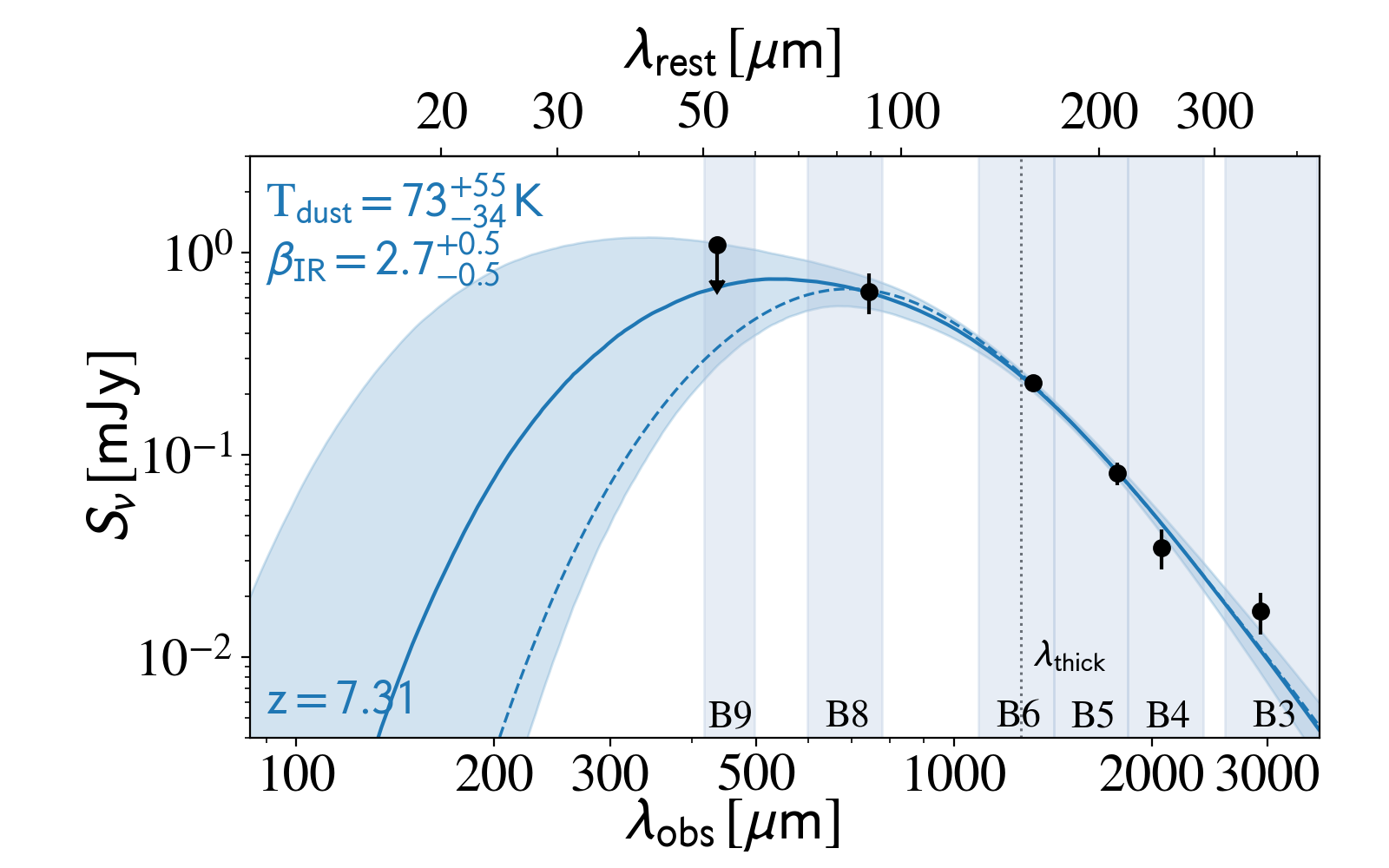}
    \caption{Optically thick modified blackbody fits to the six-band ALMA observations of REBELS-25. \textbf{Left:} Fit adopting a fixed $\lambda_\mathrm{thick} = 65\,\mu$m. \textbf{Right:} Fit allowing $\lambda_\mathrm{thick}$ to vary. In both panels, the dotted grey line indicates the wavelength where $\tau_\nu(\lambda_\mathrm{thick}) = 1$. In the fit where $\lambda_\mathrm{thick}$ is allowed to vary, the SED is poorly constrained at short wavelengths, resulting in large uncertainties on the dust temperature, mass and IR luminosity. The dust emissivity index, on the other hand, steepens slightly, but remains well-constrained.}
    \label{fig:MBB_thick}
\end{figure*}

In Section \ref{sec:results} we fit the dust SED of REBELS-25 with two optically thick models: one with a fixed thick-to-thin transition wavelength $\lambda_\mathrm{thick} = 65\,\mu$m (Sec.\ \ref{sec:results_thick}), and one where this wavelength is allowed to freely vary (Sec.\ \ref{sec:results_vary_thick}). The two fits are shown in Fig.\ \ref{fig:MBB_thick}, and the corresponding fitting parameters are presented in Table \ref{tab:fluxes}.

\section{MBB-Fitting of a Mock Galaxy at Redshift 7}
\label{app:mock}

\subsection{A Comparison of Multi-band ALMA setups}
\label{app:mock_comparison}

We are interested in finding an optimal observing strategy to constrain the dust mass, temperature, emissivity index and infrared luminosity for the $z\sim6 - 7.5$ galaxy population with ALMA. To this end, we fit mock ALMA photometry of a hypothetical high-redshift galaxy to investigate the expected constraints on its dust parameters. As outlined in Section \ref{sec:discussion_mockfitting}, our starting point is a mock $z=7$ galaxy with prior dust continuum measurements in ALMA Bands 6 and 8.

The first question is then: given observations in Bands 6 and 8, is it better to improve constraints near the peak of the dust SED with ALMA Band 9, or to focus on the Rayleigh-Jeans tail with Bands 3 or 4? \citet{bakx2021} demonstrated Band 9 to be important in order to more tightly constrain the dust temperature of galaxies with relatively warm dust ($T_\mathrm{dust} \gtrsim 45\,$K at $z\approx7$), where Band 8 is insufficient to probe the peak of the SED. On the other hand, Bands 3 or 4 are the natural choices when aiming to constrain the Rayleigh-Jeans tail of the dust SED, as they provide a longer lever arm than Band 5.

\begin{figure*}
    \centering
    \includegraphics[width=0.495\textwidth]{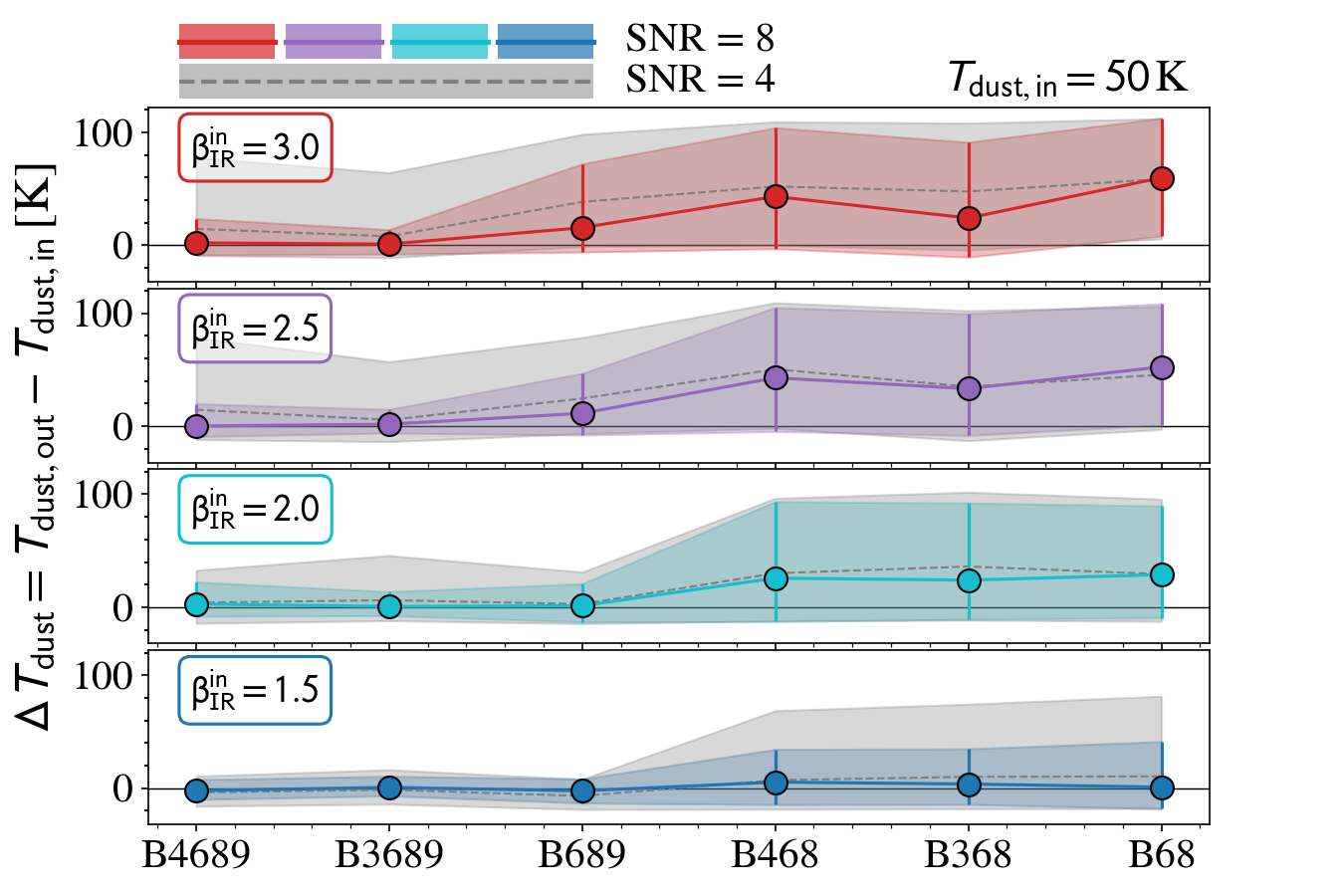} \hfill
    \includegraphics[width=0.495\textwidth]{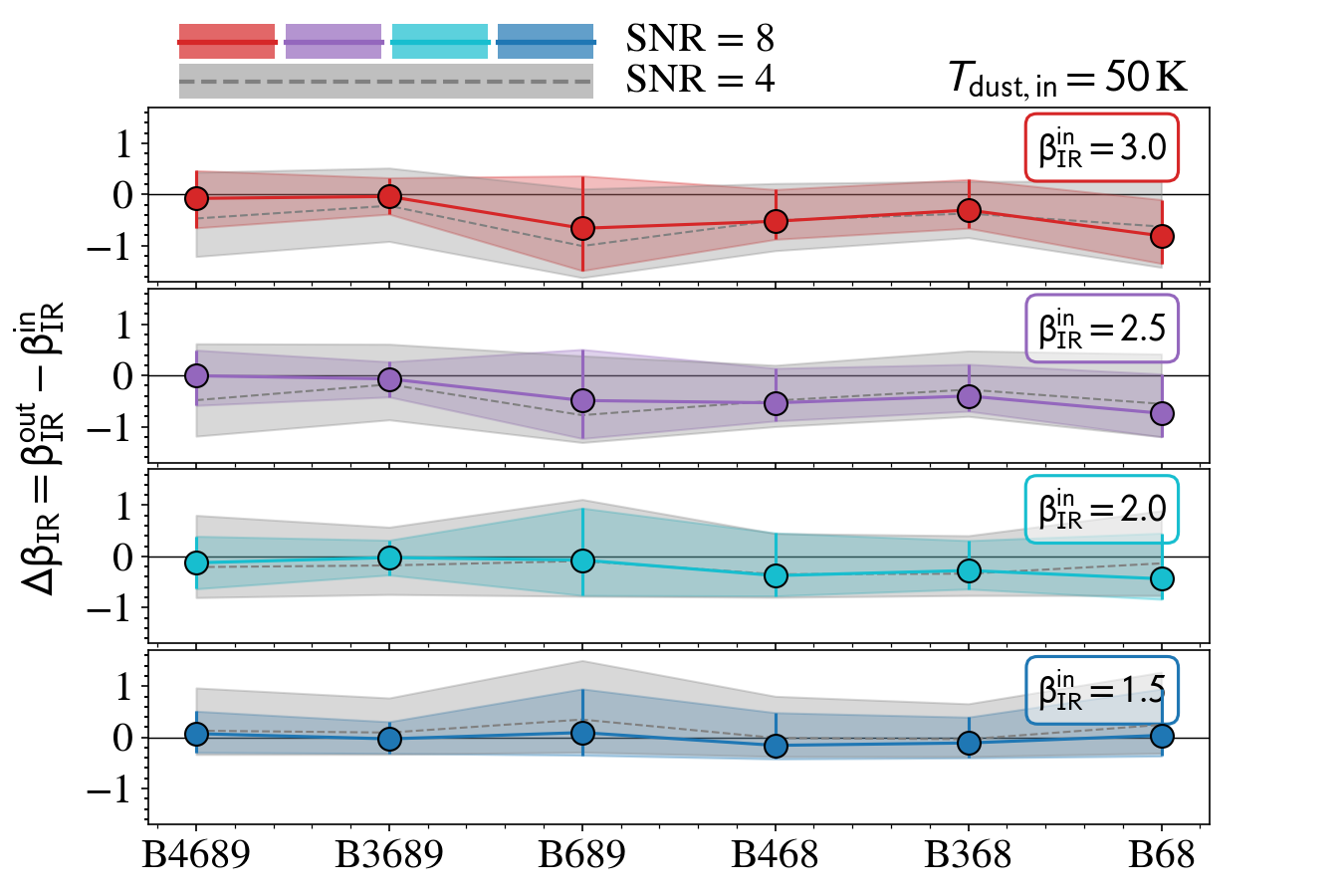} %
    \includegraphics[width=0.495\textwidth]{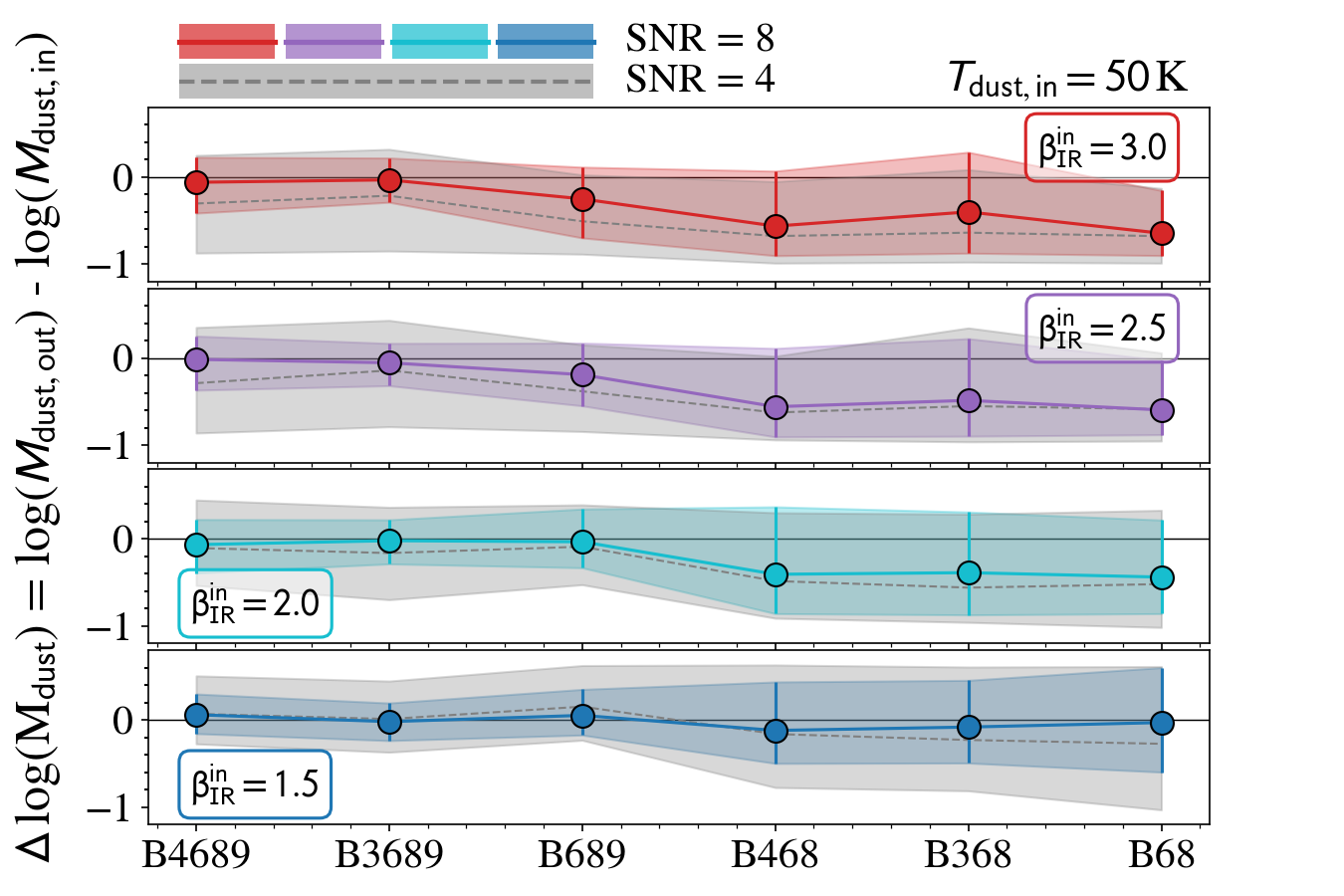} \hfill
    \includegraphics[width=0.495\textwidth]{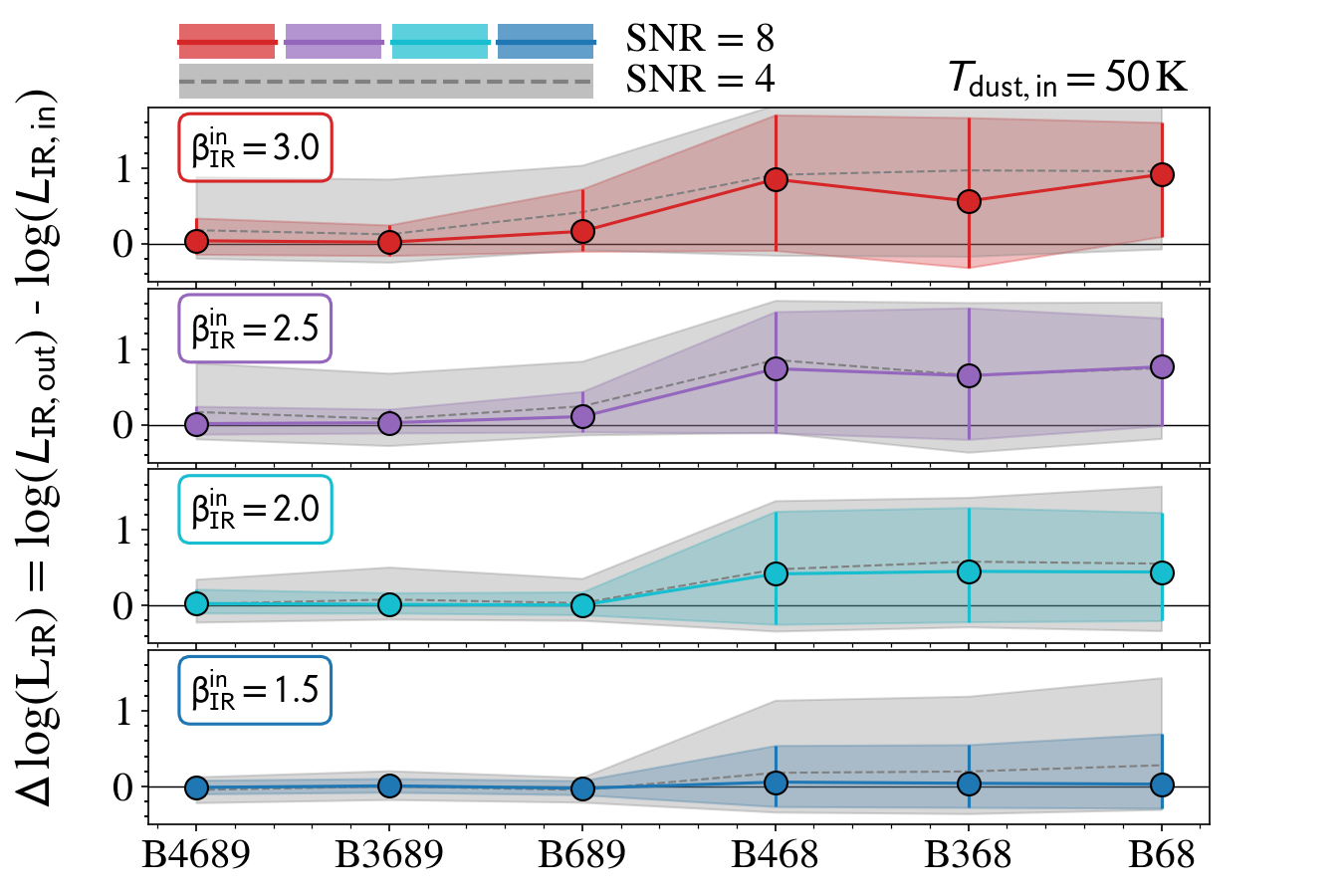}%
    \caption{Accuracy with which the key dust parameters can be constrained as a function of ALMA setup. The x-axis represents the set of ALMA bands used; for example B4689 represents Bands 4, 6, 8 and 9. \textbf{Top left:} Dust temperature. \textbf{Top right:} Dust emissivity index. \textbf{Lower left:} Dust mass. \textbf{Lower right:} Infrared luminosity. We adopt a fixed input dust temperature of $T_\mathrm{dust} = 50\,$K, and vary the input dust emissivity index between $1.5 \leq \beta_\mathrm{IR} \leq 3.0$ across the different rows. The colored shading and circular markers represent the median and $1\sigma$ confidence interval when each of the bands is assigned a S/N = 8. The dashed grey line and grey shading represent the median and corresponding uncertainty when S/N = 4.}
    \label{fig:mockfitting_setup}
\end{figure*}

To investigate which of these observing setups is most efficient, we simulate a set of optically thin modified blackbodies across a grid of dust temperatures ($T_\mathrm{dust} = 30 - 75\,$K in steps of $5\,$K) and emissivity indices ($\beta_\mathrm{IR} = [1.5, 2.0, 2.5, 3.0]$). We normalize the SED to $200\,\mu$Jy at $\nu_\mathrm{obs} = 220\,$GHz -- corresponding to a REBELS-25-like galaxy at ALMA Band 6 -- which implicitly sets the dust mass. The exact normalization is arbitrary, as the accuracy of the fitting depends on the assigned S/N, and not on the flux density. We compare six setups consisting of ALMA Bands [6, 8], [3, 6, 8], [4, 6, 8], [6, 8, 9], [3, 6, 8, 9] and [4, 6, 8, 9], in a `pessimistic' scenario where each of the bands is assigned S/N = 4, and an `optimistic' scenario where we adopt S/N = 8. We add a randomly drawn Gaussian noise realization to each flux density based on the assigned S/N, and repeat this process 50 times for each input MBB. We fit all combinations of MBBs to assess how accurately we recover the input parameters, and how important the different ALMA bands are in constraining them. We show the results for an input dust temperature of $T_\mathrm{dust,in} = 50\,$K in Fig.\ \ref{fig:mockfitting_setup}, noting that the results are qualitatively similar when a different dust temperature within the explored range of $30 - 75\,$K is adopted. \\

We start by discussing the two-band setup, consisting of just Bands 6 and 8. We find that, unless the dust emissivity index is shallow ($\beta_\mathrm{IR} = 1.5$) and the S/N is high ($\mathrm{S/N} = 8$), inferred dust temperatures and IR luminosities may be significantly overestimated, while dust masses are underestimated. This is not unexpected, given that we fit three free parameters -- each of which has rather wide priors -- to two datapoints. However, it showcases once more how two bands are not sufficient to accurately constrain galaxy dust properties.

We therefore proceed by discussing three-band ALMA setups. Figure \ref{fig:mockfitting_setup} shows that Band 9 is crucial to accurately constrain the dust temperature, and therefore $M_\mathrm{dust}$ and $L_\mathrm{IR}$, in agreement with \citet{bakx2021}. If a setup with Band 3 or 4 is adopted in favor of one with Band 9, $\beta_\mathrm{IR}$ can be slightly more accurately constrained. However, for most purposes this is unlikely to outweigh the resulting lack of constraints on the dust temperature and mass. As expected, the longer lever arm provided by Band 3 results in marginally better constraints when employing a setup consisting of Bands [3, 6, 8] compared to [4, 6, 8], assuming a fixed S/N. However, in practice, detecting dust continuum emission is much easier in Band 4, and as such it is likely to be the more efficient choice when accounting for (the lack of) available observing time.\footnote{The thermal noise level reached in a fixed observing time is similar between ALMA Bands 3 and 4, while a galaxy with $\beta_\mathrm{IR} = 2$ is expected to be roughly 4 times brighter in Band 4. As a result, dust continuum observations are more efficient in Band 4.}

With a four-band setup, including either Bands 3 or 4, in addition to Bands 6, 8 and 9, we can accurately recover the dust mass, temperature and emissivity index provided the S/N is sufficiently high. We here recommend a setup of Bands [4, 6, 8, 9] given the balance between the decreasing flux towards longer wavelengths and the sensitivity of the ALMA bands.

\begin{figure*}
    \centering 
    \includegraphics[width=1.0\textwidth]{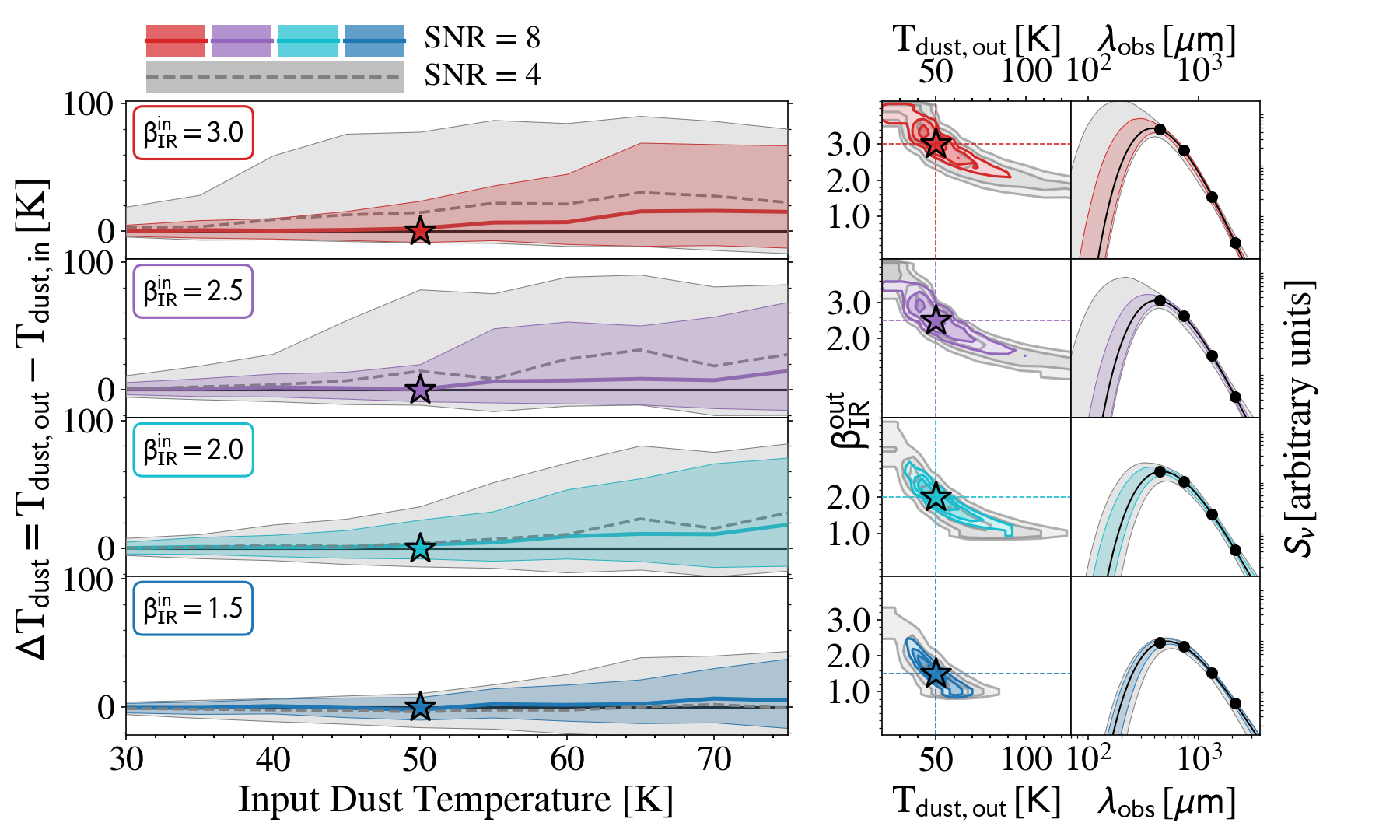}
    \caption{Same as Figure \ref{fig:mock_fitting}, now comparing a single ALMA setup of Bands [4, 6, 8, 9] at S/N = 8 (color) and S/N = 4 (grey). A high S/N is necessary to accurately constrain the dust properties of galaxies with dust emissivity indices that are steeper than the canonical $\beta_\mathrm{IR} \approx 1.5 - 2$.}
    \label{fig:mock_fitting_lowsnr}
\end{figure*}

\subsection{Bands 4, 6, 8 and 9 at S/N = 4}
\label{app:mock_lowsnr}

In Section \ref{sec:mockfitting}, we investigate to what extent adding ALMA Band 4 to a three-band setup consisting of Bands 6, 8 and 9 improves the expected constraints on the dust SED of a mock $z = 7$ galaxy. Figure \ref{fig:mock_fitting} in the main text shows the results for a fiducial S/N = 8 for all bands, while Figure \ref{fig:mock_fitting_lowsnr} shows the results for a lower S/N = 4. As expected, the simulated constraints on the dust SED are significantly poorer given this lower signal-to-noise ratio, despite utilizing a four-band setup.


\bsp	
\label{lastpage}
\end{document}